\documentclass[]{aa}

\def\OutputDriver{pdftex}

\usepackage{amsmath}
\usepackage{amssymb}
\usepackage{amsfonts}
\usepackage{bm}
\usepackage{eucal}

\usepackage[\OutputDriver]{graphicx}
\usepackage{natbib}

\usepackage{color}

\usepackage[\OutputDriver,hyperindex]{hyperref}
\hypersetup{
breaklinks = {true},
colorlinks = {false},
linkcolor={black},
pdfpagemode = {None}, 
pdfborder = {0 0 1},
pdftitle = {Ray-tracing through the Millennium Simulation: Born corrections and lens-lens coupling in cosmic shear and galaxy-galaxy lensing},
pdfsubject = {},
pdfauthor = {S. Hilbert, J. Hartlap, S.D.M. White, \& P. Schneider},
pdfkeywords = {gravitational lensing, dark matter, large-scale structure of the Universe, galaxies: general, cosmology: theory, methods: numerical}
}

\bibpunct{(}{)}{;}{a}{}{,}

\newcommand{\mrm}[1]{\mathrm{#1}}

\newcommand{\vect}[1]{\boldsymbol{#1}}

\newcommand{\parder}[3][]{\frac{\partial^{#1} {#2}}{\partial {#3}^{#1}}}

\newcommand{\p}{\partial}
\newcommand{\e}{\mathrm{e}}
\newcommand{\diff}[2][]{\mrm{d}^{#1}{#2}\,}
\newcommand{\EV}[1]{\left\langle{#1}\right\rangle}

\newcommand{\Z}{\mathbb{Z}}

\newcommand{\kpc}{\ensuremath{\mathrm{kpc}}}
\newcommand{\Mpc}{\ensuremath{\mathrm{Mpc}}}
\newcommand{\Gpc}{\ensuremath{\mathrm{Gpc}}}

\newcommand{\Msolar}{\ensuremath{\mathrm{M}_\odot}}

\newcommand{\arcsect}{\ensuremath{\mathrm{arcsec}}}
\newcommand{\arcmint}{\ensuremath{\mathrm{arcmin}}}
\newcommand{\degt}{\ensuremath{\mathrm{deg}}}

\newcommand{\zS}{z^\mrm{S}}

\newcommand{\clight}{\ensuremath{\mathrm{c}}}

\begin{document}

\title{Ray-tracing through the Millennium Simulation:\\Born corrections and lens-lens coupling in cosmic shear and galaxy-galaxy lensing}
\titlerunning{Born corrections and lens-lens coupling in cosmic shear and galaxy-galaxy lensing}
\author{S.\ Hilbert\inst{1,2}\thanks{\texttt{shilbert@astro.uni-bonn.de}} \and J.\ Hartlap\inst{2} \and S.D.M.\ White\inst{1} \and P.\ Schneider\inst{2}}

\institute{
Max Planck Institute for Astrophysics, Karl-Schwarzschild-Str. 1, 85741 Garching, Germany
\and Argelander-Institut f\"ur Astronomie, Universit\"at Bonn, Auf dem H\"ugel 71, 53121 Bonn, Germany
}

\date{Received  / Accepted }

\keywords{gravitational lensing -- dark matter -- large-scale structure of the Universe -- cosmology: theory -- methods: numerical}

\abstract{
Weak-lensing surveys need accurate theoretical predictions for interpretation of their results and cosmological-parameter estimation.
}{
We study the accuracy of various approximations to cosmic shear and weak galaxy-galaxy lensing and investigate effects of Born corrections and lens-lens coupling.
}{
We use ray-tracing through the Millennium Simulation, a large $N$-body simulation of cosmic structure formation, to calculate various cosmic-shear and galaxy-galaxy-lensing statistics. We compare the results from ray-tracing to semi-analytic predictions.
}{
(i) We confirm that the first-order approximation (i.e. neglecting lensing effects beyond first order in density fluctuations) provides an excellent fit to cosmic-shear power spectra as long as the actual matter power spectrum is used as input.
Common fitting formulae, however, strongly underestimate the cosmic-shear power spectra (by $>30\%$ on scales $\ell>10000$). 
Halo models provide a better fit to cosmic shear-power spectra, but there are still noticeable deviations ($\sim10\%$).
(ii) Cosmic-shear B-modes, which are induced by Born corrections and lens-lens coupling, are at least three orders of magnitude smaller than cosmic-shear E-modes. Semi-analytic extensions to the first-order approximation predict the right order of magnitude for the B-mode. Compared to the ray-tracing results, however, the semi-analytic predictions may differ by a factor two on small scales and also show a different scale dependence.
(iii) The first-order approximation may under- or overestimate the galaxy-galaxy-lensing shear signal by several percent due to the neglect of magnification bias, which may lead to a correlation between the shear and the observed number density of lenses.
}{
(i) Current semi-analytic models need to be improved in order to match the degree of statistical accuracy expected for future weak-lensing surveys.
(ii) Shear B-modes induced by corrections to the first-order approximation are not important for future cosmic-shear surveys.
(iii) Magnification bias can be important for galaxy-galaxy-lensing surveys.
}

\maketitle

\section{Introduction}

During the past few years, weak gravitational lensing has developed rapidly from mere detection to an important cosmological tool \citep{MunshiEtal2008_wl_review}.
Measurements of cosmic shear help us to constrain the properties of the cosmic matter distribution \citep[e.g.][]{SemboloniEtal2006,HoekstraEtal2006,SimonEtal2007,BenjaminEtal2007,MasseyEtal2007_cosmic_scaffolding,FuEtal2008}, the growth of structure \citep[e.g.][]{BaconEtal2005,MasseyEtal2007_3D_WL}, and the nature of the dark energy \citep[e.g.][]{TaylorEtal2007,SchimdEtal2007,AmendolaKunzSapone2008}.
Weak galaxy-galaxy lensing can be used to study the properties of galactic dark-matter halos and the relation between luminous and dark matter \citep[e.g.][]{MandelbaumEtal2006_Galaxies,SimonEtal2007,GavazziEtal2007}.

The accuracy that can be reached in weak-lensing surveys is determined by several factors. On the observational side, high accuracy requires large field sizes and deep observations with a high number density of galaxies with measurable shapes. Moreover, it is crucial to obtain an accurate and unbiased measurement of galaxy ellipticities. Finally, for the interpretation of the resulting data and the inference of cosmological parameters, an accurate theoretical model is needed.
A thorough understanding of systematic effects in weak lensing will become particularly important with the advent of very large weak-lensing surveys such as CFHTLS\footnote{\texttt{http://www.cfht.hawaii.edu/Science/CFHLS}}, KIDS\footnote{\texttt{http://http://www.astro-wise.org/projects/KIDS}}, Pan-STARRS\footnote{\texttt{http://pan-starrs.ifa.hawaii.edu}}, and LSST\footnote{\texttt{http://www.lsst.org}}, or the planned Dark Energy Survey\footnote{\texttt{http://www.darkenergysurvey.org}}, DUNE \citep{RefregierEtal2006_DUNE}, and SNAP\footnote{\texttt{http://snap.lbl.gov}}. For these surveys, the statistical uncertainties will be very small, so the accuracy will be limited by the remaining systematics in the data reduction and theoretical modeling.

While significant improvement on image-ellipticity measurements are expected in the near future \citep{MasseyEtal2007_STEP2}, one still needs to investigate, how uncertain current theoretical predictions are, and how much improvement can be expected for these. Presently, the most accurate way to obtain predictions for weak-lensing surveys is to perform ray-tracing through large high-resolution $N$-body simulations of cosmic structure formation \citep[see, e.g.,][]{WambsganssCenOstriker1998,JainSeljakWhite2000,WhiteHu2000,VanWaerbekeEtal2001,HamanaMellier2001,ValeWhite2003,White2005}. The drawback of this approach is that large $N$-body simulations are computationally demanding, so using them to explore the whole parameter space of cosmological models is currently unrealistic. On the other hand, ray-tracing simulations enable one to check the approximations and assumptions made in computationally less demanding (semi-)analytic models, and adjust and extend these models where necessary.

Numerous ray-tracing methods have been developed to study the many aspects of gravitational lensing. Tree-based ray-tracing methods \citep[][]{AubertAmaraMetcalf2007} that adapt to the varying spatial resolution of $N$-body simulations have been used to study the impact of substructure on strong lensing by dark matter halos \citep[][]{PeiraniEtal2008}. Cluster strong lensing simulations, which require good mass modelling of galaxy clusters, usually ignore the matter distribution outside clusters and use the thin-lens approximation in the ray-tracing \citep[e.g.][]{BartelmannWeiss1994,MeneghettiEtal2007,RozoEtal2008}.

Many simulations of weak lensing by clusters and large-scale structure \citep[e.g.][]{WambsganssCenOstriker1998,JainSeljakWhite2000,ValeWhite2003,PaceEtal2007} employ algorithms that are based on the multiple-lens-plane approximation \citep{BlandfordNarayan1986} to trace light rays through cosmological $N$-body simulations. Others \citep[e.g.][]{CouchamnBarberThomas1999,CarboneEtal2008_I} perform ray-tracing though the three-dimensional gravitational potential. In a simpler approach \citep[e.g.][]{WhiteVale2004,HeymansEtal2006,HilbertMetcalfWhite2007}, the matter in the $N$-body simulation is projected along unperturbed light paths onto a single lens plane, which is then used to calculate lensing observables. Recent simulations of CMB lensing use generalisations of the single- or multiple-plane approximation that take the curvature of the sky into account \citep[e.g.][]{DasBode2008,TeyssierEtal2008_arXiv,FosalbaEtal2008}.

In this work, we employ multiple-lens-plane ray-tracing through the Millennium Simulation \citep{SpringelEtal2005_Millennium} to study weak lensing. One of the largest $N$-body simulations available today, the Millennium Simulation provides not only a much larger volume, but also a higher spatial and mass resolution than simulations used for earlier weak-lensing studies. In order to take full advantage of the large simulation volume and high resolution, the ray-tracing algorithm used here differs in several aspects from algorithms used in previous works \citep[e.g.][]{JainSeljakWhite2000}. Here, we give a detailed description of our ray-tracing algorithm.

Semi-analytic weak-lensing predictions are usually based on the first-order approximation, in which light deflections are only considered to first order in the peculiar gravitational potential and hence, to first order in the matter fluctuations. The ray-tracing approach allows us to look at effects neglected in the first-order approximation such Born corrections and lens-lens coupling. Here, we investigate the cosmic-shear B-modes induced by these effects and compare the ray-tracing results to semi-analytic estimates \citep{CoorayHu2002,HirataSeljak2003,ShapiroCooray2006}, whose accuracy has not been confirmed by numerical simulations yet. Moreover, we investigate how well fitting formulae \citep{PeacockDodds1996,EisensteinHu1999,SmithEtal2003} and halo models \citep{Seljak2000,CooraySheth2002} reproduce cosmic-shear power spectra. Finally, we investigate the accuracy of the first-order approximation for weak galaxy-galaxy lensing.

The paper is organised as follows. In the next section, we introduce the theoretical background and notation used in our lensing analysis. In Sec.~\ref{sec:algorithm}, we discuss our ray-tracing algorithm. The results from our ray-tracing analysis are presented in Sec.~\ref{sec:results}. We conclude our paper with a summary in Sec.~\ref{sec:summary}.

\section{Theory}
\label{sec:theory}

\subsection{Gravitational light deflection}
\label{sec:light_deflection}

In this section, we introduce the formulae relating the `apparent' positions of distant light sources to their `true' positions.  In order to label spacetime points in a model universe with a weakly perturbed Friedmann-Lema{\^i}tre-Robertson-Walker (FLRW) metric, we choose a coordinate system $(t,\vec{\beta},w)$ based on physical time $t$, two angular coordinates $\vec{\beta}=(\beta_1,\beta_2)$, and the line-of-sight comoving distance $w$ relative to the observer. The spacetime metric of the model universe is then given by \citep[see, e.g.,][]{BartelmannSchneider2001_WL_review}:
\begin{multline}
\label{eq:perturbed_FLRW_metric}
\diff{s^2}=
\left(1+\frac{2\Phi}{\clight^2}\right)\clight^2\diff{t^2} - \left(1-\frac{2\Phi}{\clight^2}\right)a^2
\\\times
\bigg\{\diff{w^2}+f_K^2(w)\left[\diff{\beta_1^2}+ \cos^2(\beta_1)\diff\beta{_2^2}\right]\bigg\}.
\end{multline}
Here, $\clight$ denotes the speed of light, $a=a(t)$ denotes the scale factor, and $\Phi=\Phi(t,\vec{\beta},w)$ denotes the peculiar (Newtonian) gravitational potential. The comoving angular diameter distance is defined as:
\begin{equation}
\label{chap:Gravitational_lensing:eq:def_f_K}
f_K(w)= \begin{cases}
1/\sqrt{K}\sin\left(\sqrt{K}w\right) & \text{for } K>0,\\
w &  \text{for } K=0,\text{ and}\\
1/\sqrt{-K}\sinh\left(\sqrt{-K}w\right) & \text{for } K<0,
\end{cases}
\end{equation}
where $K$ denotes the curvature of space. The particular choice for the angular coordinates $\vec{\beta}=(\beta_1,\beta_2)$ is convenient for the application of the `flat-sky' approximation, where the metric near $\vect{\beta}=\vect{0}$ is approximated using $\cos^2(\beta_1)\approx1$.

Consider the path, parametrised by comoving distance $w$,  of a photon eventually reaching the observer from angular direction $\vec{\theta}$. The angular position $\vec{\beta}(\vec{\theta},w)$ of the photon at comoving distance $w$ is then given by \citep[see, e.g.,][for a sketch of a derivation]{JainSeljak1997}:
\begin{multline}
\label{eq:cont_lens_eq}
\vec{\beta}(\vec{\theta},w)  =
 \vec{\theta}  - \frac{2}{\clight^2} \int_0^w \!\!\diff{w'} \frac{f_K(w-w')}{f_K(w)f_K(w')}
\\\times
\nabla_{\vec{\beta}} \Phi \big(t(w'),\vec{\beta}(\vec{\theta},w'),w'\big)
\end{multline}
with $\nabla_{\vec{\beta}}=(\p/\p\beta_1,\p/\p \beta_2)$, and $t(w')$ denoting the cosmic time of events at line-of-sight comoving distance $w'$ from the observer. By differentiation this equation  w.r.t.  $\vec{\theta}$, we obtain the distortion matrix $\tens{A}$, i.e. the Jacobian of the lens mapping $\vec{\theta}\mapsto \vec{\beta}=\vec{\beta}(\vec{\theta},w)$:
\begin{equation}
  \label{eq:cont_Jacobian}
 \begin{split}
  \tens{A}_{ij}(\vec{\theta},w) &=
\parder{\beta_i(\vec{\theta},w)}{\theta_j}
\\&=
\delta_{ij} - \frac{2}{c^2}\int_{0}^w \diff{w'}\frac{f_K(w-w')}{f_K(w)f_K(w')}
\\&\qquad\times
\parder{^2\Phi \big(t(w'),\vec{\beta}(\vec{\theta},w'),w'\big)}{\beta_i\partial\beta_k} \tens{A}_{kj}(\vec{\theta},w').
\end{split}
\end{equation}

Due to the matrix products in Eq.~\eqref{eq:cont_Jacobian}, the distortion matrix $\tens{A}$ is generally not symmetric. However, it can be decomposed into a rotation matrix (related to a usually unobservable rotation in the source plane) and a symmetric matrix \citep{SchneiderEhlersFalco_book}:
\begin{multline}
\label{eq:distortion_standard_decomposition}
\tens{A}(\vect{\theta},w)=
\begin{pmatrix}
\cos\omega & \sin\omega\\
-\sin\omega &\cos\omega
\end{pmatrix}
\\\times
\begin{pmatrix}
1-\kappa-\gamma_1 & -\gamma_2\\
-\gamma_2 & 1-\kappa+\gamma_1
\end{pmatrix}.
\end{multline}
The decomposition defines the rotation angle $\omega=\omega(\vect{\theta},w)$, the convergence $\kappa=\kappa(\vect{\theta},w)$, and the two components $\gamma_1=\gamma_1(\vect{\theta},w)$ and $\gamma_2=\gamma_2(\vect{\theta},w)$ of the shear, which may be combined into the complex shear $\gamma=\gamma_1+\mrm{i} \gamma_2$.

The shear field $\gamma(\vect{\theta},w)$ can be decomposed into a rotation-free part $\gamma_\mrm{E}(\vect{\theta},w)$ and a divergence-free part $\gamma_\mrm{B}(\vect{\theta},w)$. For infinite fields, the decomposition into these \emph{E/B-modes} is most easily written down in Fourier space:
\begin{subequations}
\label{eq:EB_fourier_decomposition}
\begin{align}
\hat{\gamma}_\mrm{E}(\vect{\ell},w) &= \frac{\ell^2}{|\ell|^4}\left[(\ell_1^2-\ell_2^2)\hat{\gamma}_1(\vect{\ell},w) + 2\ell_1\ell_2\hat{\gamma}_2(\vect{\ell},w)  \right]
,\\
\hat{\gamma}_\mrm{B}(\vect{\ell},w) &= \frac{\ell^2}{|\ell|^4}\left[(\ell_1^2-\ell_2^2)\hat{\gamma}_2(\vect{\ell},w) - 2\ell_1\ell_2\hat{\gamma}_1(\vect{\ell},w)   \right]
.
\end{align}
\end{subequations}
Here, hats denote Fourier transforms, $\vect{\ell}=(\ell_1,\ell_2)$ denotes the Fourier wave vector, and $\ell=\ell_1+\mrm{i}\ell_2$. Care must be taken when decomposing the shear in fields of finite size, where the field boundaries can cause artifacts \citep{SeitzSchneider1996}. These artifacts can be avoided by using aperture masses to quantify the shear E- and B-mode contributions \citep{CrittendenEtal2002,SchneiderVanWaerbekeMellier2002}.

Equations \eqref{eq:cont_lens_eq} and \eqref{eq:cont_Jacobian} are implicit relations for the light path and the Jacobian. The solution of  Eq.~\eqref{eq:cont_lens_eq} to first order in the potential is obtained by integrating along undisturbed light paths:
\begin{multline}
  \label{eq:defl_LinearApprox}
\vec{\beta}(\vec{\theta},w)  =
 \vec{\theta}  - \frac{2}{\clight^2} \int_0^w \!\!\diff{w'} \frac{f_K(w-w')}{f_K(w)f_K(w')}
\\\times
\nabla_{\vec{\theta}} \Phi \big(t(w'),\vec{\theta},w'\big)
.
\end{multline}
The distortion to first order reads:
\begin{multline}
  \label{eq:cont_Jacobian_LinearApprox}
  \tens{A}_{ij}(\vec{\theta},w) =
\delta_{ij} - \frac{2}{c^2}\int_{0}^w \diff{w'}\frac{f_K(w-w')}{f_K(w)f_K(w')}
\\\times
\parder{^2\Phi \big(t(w'),\vec{\theta},w'\big)}{\theta_i\partial\theta_k}.
\end{multline}
The first-order approximation to the distortion contains the \emph{Born approximation}, which ignores deviations of the actual light path from the undisturbed path on the r.h.s. of Eq.~\eqref{eq:cont_Jacobian}. Moreover, \emph{lens-lens coupling} is neglected, i.e.\ the appearance of the distortion on the r.h.s.\ of Eq.~\eqref{eq:cont_Jacobian}.
The neglected lens-lens coupling and corrections to the Born approximation account for the effect that light from a distant source `sees' a distorted image of the lower-redshift matter distribution due to higher-redshift matter inhomogeneities along the line-of-sight. Thus, the first-order approximation works well in regions where larger matter inhomogeneities are absent or confined to a small redshift range, but fails in regions where noticeable distortions arise from matter inhomogeneities at multiple redshifts.

Born corrections and lens-lens coupling effects may create shear B-modes. The perturbative calculation of the shear B-modes by iteratively solving Eq.~\eqref{eq:cont_Jacobian} is possible~\citep{CoorayHu2002,HirataSeljak2003}, but tedious, and the accuracy of this approach is not known. However, multiple deflections and lens-lens coupling effects are fully included in the multiple-lens-plane approximation as described below. We will thus use this approximation to investigate these effects and assess the quality of perturbative calculations of these effects.

\subsection{The multiple-lens-plane approximation}
\label{sec:MLPA}

In the multiple-lens-plane approximation~\citep[see, e.g.,][]{BlandfordNarayan1986,SchneiderEhlersFalco_book,SeitzSchneiderEhlers1994,JainSeljakWhite2000}, a series of lens planes perpendicular to the central line-of-sight is introduced into the observer's backward light cone.  The continuous deflection that a light ray experiences while propagating through the matter inhomogeneities in the light cone is then approximated by finite deflections at the lens planes. The deflections are calculated from a projected matter distribution on the lens planes. This corresponds to solving the integral equations \eqref{eq:cont_lens_eq} and \eqref{eq:cont_Jacobian} by discretisation (and using the impulse approximation).

The deflection $\vec{\alpha}^{(k)}(\vec{\beta}^{(k)})$ of a light ray intersecting the $k^\mrm{th}$ lens plane (here, we count from the observer to the source) at angular position $\vec{\beta}^{(k)}$ can be expressed as the gradient of a lensing potential $\psi^{(k)}$:
\begin{equation}
\label{eq:deflangle}
{\vec{\alpha}}^{(k)}(\vec{\beta}^{(k)})=\vec{\nabla}_{\vec{\beta}^{(k)}}\,\psi^{(k)}(\vec{\beta}^{(k)})\; .
\end{equation}
The differential deflection is then given by higher derivatives of the lensing potential. The second derivatives can be combined into the shear matrix
\begin{equation}
\tens{U}^{(k)}_{ij}=\frac{\partial^2\psi^{(k)}(\vec{\beta}^{(k)})}{\partial \beta^{(k)}_i\partial \beta^{(k)}_j} = \frac{\partial\alpha^{(k)}_i(\vec{\beta}^{(k)})}{\partial \beta^{(k)}_j}.
\end{equation}
The lensing potential  $\psi^{(k)}$ is a solution of the Poisson equation:
\begin{equation}
\vec{\nabla}_{\vec{\beta}^{(k)}}^2\psi^{(k)}(\vec{\beta}^{(k)}) = 2 \sigma^{(k)}(\vec{\beta}^{(k)}).
\end{equation}
The dimensionsless surface mass density $\sigma^{(k)}$ is given by a projection of the matter distribution in a slice around lens plane:
\begin{equation}
\sigma^{(k)}(\vec{\beta}^{(k)}) =
\frac{3H_0^2\Omega_{\rm m}}{2c^2}\,\frac{f_K^{(k)}}{a^{(k)}}
\int_{w^{(k)}_{\mrm{L}}}^{w^{(k)}_{\mrm{U}}} \diff{w'}\delta_\text{m}\big(\vec{\beta}^{(k)},w'\big).
\end{equation}
Here, $H_0$ denotes the Hubble constant, $\Omega_\text{m}$ the mean matter density in terms of the critical density, $f_K^{(k)}=f_K(w^{(k)})$ and $a^{(k)}=a(w^{(k)})$, with $w^{(k)}$ denoting the 
line-of-sight comoving distance of the plane. Furthermore, $\delta_\text{m}\big(\vec{\beta}^{(k)},w'\big)$ denotes the three-dimensional density contrast at comoving
position $\big(\vec{\beta}^{(k)},w'\big)$ relative to the mean matter density. The slice boundaries $w^{(k)}_{\mrm{L}}$ and
$w^{(k)}_{\mrm{U}}$ have to satisfy $w^{(k)}_{\mrm{L}}<w^{(k)}<w^{(k)}_{\mrm{U}}$ and $w^{(k)}_{\mrm{U}}=w^{(k+1)}_{\mrm{L}}$. They are 
usually chosen to correspond to the mean redshifts~\citep[e.g.][]{JainSeljakWhite2000} or comoving distances~\citep[e.g.][]{WambsganssBodeOstriker2004} of successive 
planes.\footnote{The exact choice for the projection boundaries becomes unimportant for sufficiently small spacings between the lens planes.}
These conditions ensure that every region of the light cone contributes exactly to one lens plane, which is the closest plane in redshift or comoving distance.

\begin{figure*}[t]
\centerline{\includegraphics[width=0.85 \linewidth]{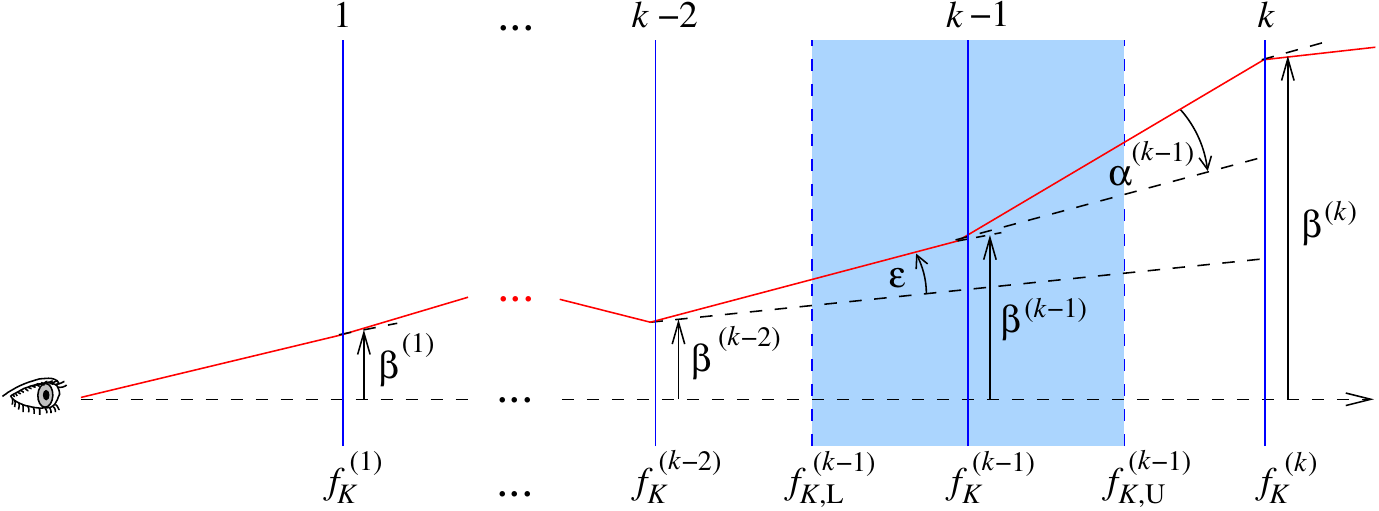}}
\caption
{
\label{fig:light_cone}
Schematic view of the observer's backward light cone in the multiple-lens-plane approximation. A light ray (red line) experiences a deflection only when passing through a lens plane (solid blue lines). The deflection angle $\vec{\alpha}^{(k-1)}$ of a ray passing through the lens plane at distance $f_{K}^{(k-1)}$ from the observer is obtained from the matter distribution between $f_{K,\mrm{U}}^{(k-1)}$ and  $f_{K,\mrm{L}}^{(k-1)}$ projected onto the plane.
Using the deflection angle $\vec{\alpha}^{(k-1)}$ of the light ray at the previous lens plane and the ray's angular positions $\vec{\beta}^{(k-1)}$ and $\vec{\beta}^{(k-2)}$ on the two previous planes, the angular position $\vec{\beta}^{(k)}$ on the current plane can be computed.
}
\end{figure*}

Given the deflection angles on the lens planes, one can trace back a light ray reaching the observer from angular position $\vec{\beta}^{(1)}=\vec{\theta}$ on the first lens plane to the other planes:
\begin{equation}
\label{eq:lens_mapping_old_style}
\vec{\beta}^{(k)}(\vec{\theta})=\vec{\theta}-\sum_{i=1}^{k-1}\frac{f_K^{(i,k)}}{f_K^{(k)}}\vec{\alpha}^{(i)}(\vec{\beta}^{(i)})
\;,\;\;k=1,2,\ldots
\end{equation}
Here, $f_K^{(i,k)}=f_K\big(w^{(k)}-w^{(i)}\big)$.

Equation \eqref{eq:lens_mapping_old_style} is not practical for tracing rays through many lens planes. An alternative expression is obtained as follows 
\citetext{see, e.g., \citealp{Hartlap2005_Diplomarbeit}, or \citealp{SeitzSchneiderEhlers1994} for a different derivation}: The angular position $\vec{\beta}^{(k)}$ of a light ray on the lens plane $k$ is related to its positions $\vec{\beta}^{(k-2)}$ and $\vec{\beta}^{(k-1)}$ on the two previous lens planes by (see Fig.~\ref{fig:light_cone}):
\begin{equation}
\begin{split}
\label{eq:comoving_pos_iter}
f_K^{(k)}\vec{\beta}^{(k)} &=
f_K^{(k)}\vec{\beta}^{(k-2)} + f_K^{(k-2,k)}\vect{\epsilon}
\\&\quad
 -  f_K^{(k-1,k)} {\vec{\alpha}}^{(k-1)}\big(\vec{\beta}^{(k-1)}\big),
\end{split}
\end{equation}
\begin{equation*}
\text{where }
\vec{\epsilon}=
\frac{f_K^{(k-1)}}{f_K^{(k-2,k-1)}}
\left(\vec{\beta}^{(k-1)}-\vec{\beta}^{(k-2)}\right).
\end{equation*}
Hence,
\begin{equation}
\begin{split}
\label{eq:ang_pos}
\vec{\beta}^{(k)} &=
\left(1-\frac{f_K^{(k-1)}}{f_K^{(k)}} \frac{f_K^{(k-2,k)}}{f_K^{(k-2,k-1)}} \right)\vec{\beta}^{(k-2)}
\\&\quad
+ \frac{f_K^{(k-1)}}{f_K^{(k)}} \frac{f_K^{(k-2,k)}}{f_K^{(k-2,k-1)}}\vec{\beta}^{(k-1)}
\\&\quad
- \frac{f_K^{(k-1,k)}}{f_K^{(k)}} {\vec{\alpha}}^{(k-1)}\big(\vec{\beta}^{(k-1)}\big).
\end{split}
\end{equation}
For a light ray reaching the observer from angular position $\vec{\theta}$ on the first lens plane, one can compute its angular position on the other lens planes by iterating Eq.~\eqref{eq:ang_pos} with initial values $\vec{\beta}^{(0)}=\vec{\beta}^{(1)}=\vec{\theta}$.

Differentiating Eq.~\eqref{eq:ang_pos} with respect to $\vec{\theta}$, we obtain a recurrence relation for the distortion matrix:
\begin{equation}
\begin{split}
\label{eq:jacob_rec}
\tens{A}^{(k)}_{ij} &=
\left(1-\frac{f_K^{(k-1)}}{f_K^{(k)}} \frac{f_K^{(k-2,k)}}{f_K^{(k-2,k-1)}} \right)\tens{A}^{(k-2)}_{ij}
\\&\quad
+ \frac{f_K^{(k-1)}}{f_K^{(k)}} \frac{f_K^{(k-2,k)}}{f_K^{(k-2,k-1)}}\tens{A}^{(k-1)}_{ij}
\\&\quad
- \frac{f_K^{(k-1,k)}}{f_K^{(k)}}
\tens{U}^{(k-1)}_{ik}\tens{A}^{(k-1)}_{kj}.
\end{split}
\end{equation}
With the knowledge of the involved distances  and shear matrices, this equation allows us to iteratively compute the distortion matrix of a light ray from the observer to any lens plane. This equation requires in practice much fewer arithmetic operations and memory than the commonly used relations~\citep[e.g. by][]{JainSeljakWhite2000} based on Eq.~\eqref{eq:lens_mapping_old_style}.

For comparison and testing, we will also use the multiple-lens-plane algorithm to calculate the distortion in the first-order approximation by:
\begin{equation}
\label{eq:jacob_multi_plane_LinearApprox}
\tens{A}^{(k)}_{ij}(\vec{\theta})=\delta_{ij}-\sum_{n=1}^{k-1}\frac{f_K^{(n,k)}}{f_K^{(k)}}\tens{U}^{(n)}_{ij}(\vec{\theta}).
\end{equation}

\section{The ray-tracing algorithm}
\label{sec:algorithm}
The methods we use for ray-tracing through $N$-body simulations to study lensing  are generally similar to those used by, e.g., \citet{JainSeljakWhite2000} or \citet{ValeWhite2003}. First, the matter distribution on the past light cone of a fiducial observer is constructed from the simulation data. Then, the past light cone is partitioned into a series of redshift slices. The content of each slice is projected onto a lens plane. Finally, the multiple-lens-plane approximation is used to trace back light rays from the observer through the series of lens planes to the sources.

The purpose of our ray-tracing algorithm is to simulate strong and weak lensing in a way that takes full advantage of the unprecedented statistical power offered by the large volume and high spatial and mass resolution of the Millennium Simulation.\footnote{
This work concentrates on weak lensing, but the algorithm is also used for strong-lensing studies \citep{HilbertEtal2007_StrongLensing, HilbertEtal2008_StrongLensing_II,FaureEtal2008_arXiv}.}
Therefore, our ray-tracing method differs in many details from previous works. Most notably, we use a multiple-mesh method and adaptive smoothing to calculate light deflections and distortions from the projected matter distribution on the lens planes. This allows us to simulate lensing on the full range of scales covered by the Millennium Simulation, ranging from strong lensing on scales $\gtrsim1\,\arcsect$ to cosmic shear on scales $\lesssim 1\,\degt$. A brief outline of our algorithms for the construction of the past light cones and the lens planes has been given in an earlier work \citep{HilbertEtal2007_StrongLensing}. Here, we extend the discussion and provide a more detailed description.

\subsection{The Millennium Simulation}
\label{sec:Millennium_Simulation}
The Millennium Simulation~\citep{SpringelEtal2005_Millennium} is a large $N$-body simulation of cosmic structure formation in a flat $\Lambda$CDM universe. The following cosmological parameters were assumed for the simulation: a matter density of $\Omega_\mrm{m}=0.25$ in units of the critical density, a cosmological constant with $\Omega_\Lambda=0.75$, a Hubble constant $h=0.73$ in units of $100\,\mrm{km}\,\mrm{s}^{-1}\Mpc^{-1}$, a spectral index $n=1$ and a normalisation parameter $\sigma_8=0.9$ for the primordial linear density power spectrum. These chosen parameters are consistant with the 2dF~\citep{CollessEtal2003_2dF_Data} and WMAP 1st-year data analysis~\citep{SpergelEtal2003_WMAP_1stYear_Data}. The simulation followed the evolution of the matter distribution in a cubic region of $L=500h^{-1}\,\Mpc$ comoving side length from redshift $z=127$ to the present
using a TreePM version of \mbox{\textsc{gadget}-2}~\citep{Springel2005_GADGET2} with $2160^3$ particles of mass $m_\mrm{p}=8.6 \times 10^8 h^{-1}\,\Msolar$ and a force softening length of $5h^{-1}\,\kpc$ comoving.

Snapshots of the simulation were stored on disk at 64 output times. These snapshots contain, among other data, the positions, SPH smoothing lengths and friend-of-friend (FoF) group data of the particles. The storage order for the particle data is based on a spatial oct-tree decomposition of the simulation cube \citep[see][for details]{Springel2005_GADGET2,SpringelEtal2005_Millennium}, which facilitates access to the particle data for small subvolumes of the simulation.

Complex physical processes of baryonic matter such as the formation and evolution of stars in galaxies has not been incorporated directly into the Millennium Simulation. However, several galaxy-formation models have been used to predict the properties of galaxies in the simulation \citep{SpringelEtal2005_Millennium,CrotonEtal2006,BowerEtal2006,DeLuciaBlaizot2007}. The ray-tracing methods presented in this paper will allow us (in future work) not only to study cosmic shear in great detail, but also to make predictions for galaxy-galaxy lensing (and related higher-order statistics) for the various galaxy-formation models.

\subsection{The construction of the matter in the backward light cones}
\label{sec:light_cones}
Even with a comoving size of $L=500 h^{-1}\,\Mpc$, the simulation box is too small to trace back light rays within one box. We therefore exploit the periodic boundary conditions of the simulation by arranging replicas of the simulation box in a simple cubic lattice with a lattice constant equal to the box size $L$ to fill space. We refrain from randomly shifting or rotating the content of the lattice cells, because the simulation box is far too large to be projected onto a single lens plane. In addition, this allows us to keep the matter distribution continuous across the cell boundaries.

\begin{figure}[tb]
\centerline{
\includegraphics[width=0.18\linewidth]{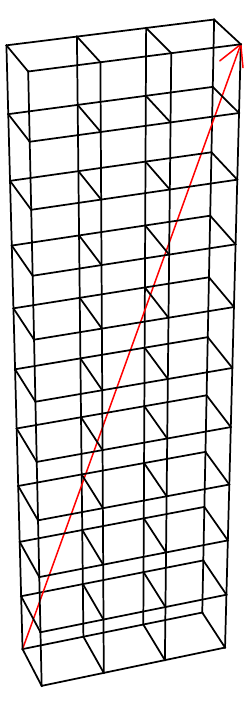}
\hspace{3em}
\includegraphics[width=0.6\linewidth]{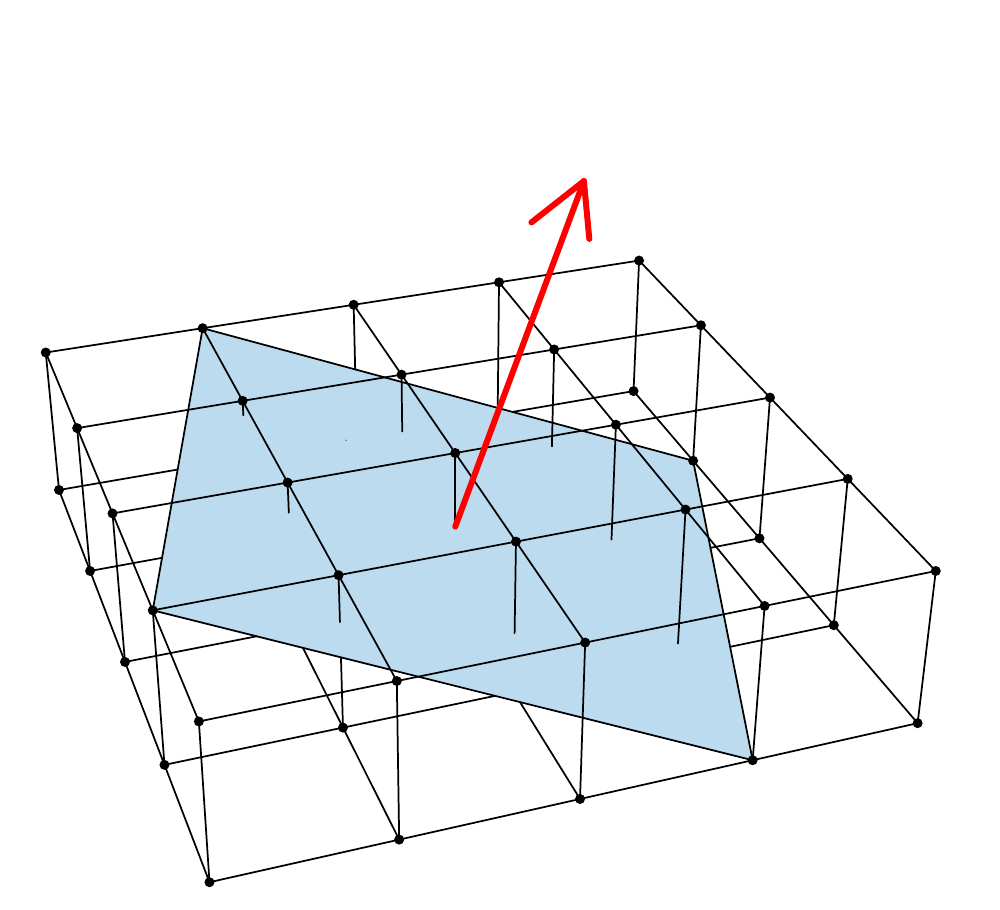}
}
\caption{
\label{fig:tilted_plane}
Schematic view of the orientation of the line-of-sight (red line) and the lens planes (blue area) relative to the simulation box (indicated by black lines).
}
\end{figure}

In this periodic matter distribution, light rays would encounter the same structures many times at different epochs before reaching relevant source redshifts if one chose the line of sight (LOS) parallel to the box edges. Hence, the LOS must be chosen at a skewed angle relative to the box axes.
On the other hand, the application of Fourier methods for the calculation of the light deflection at the lens planes requires a matter density that is periodic perpendicular to the LOS.
Choosing a LOS parallel to $\vec{n}=(n_1,n_2,n_3)$ with suitable coprime $n_i\in\Z$, one can obtain a large enough repetition length of $|\vec{n}|L$ along the LOS (see Appendix~\ref{sec:appendix_lattice_planes}). At the same time, the matter distribution is periodic perpendicular to the LOS with an area of periodicity given by $|\vec{n}|L^2$.
Our choice for $\vec{n}=(1,3,10)$ yields a LOS periodicity of $5.24h^{-1}\,\Gpc$ (corresponds to $z=3.87$) and a rectangular unit cell of $1.58h^{-1}\,\Gpc\times1.66h^{-1}\,\Gpc$ for the lens planes. Moreover, any directions with shorter periodicity are at least $1.81\deg$ away from $\vec{n}$, and a light cone with a $1.7\deg\times1.7\deg$ field of view does not intersect with itself up to redshift $z=3.87$ when folded back into the simulation cube.\footnote{We often use a larger field of view -- in particular, if only lower source redshifts are considered. Even for high source redshifts, where the resulting light cone may cover the same simulation region more than once, a large field of view can  be used with due care \citep{HilbertMetcalfWhite2007}} The resulting orientation of the LOS and the lens planes w.r.t. the simulation box are illustrated in Fig.~\ref{fig:tilted_plane}.

We partition the observer's backward light cone into redshift slices such that each slice contains the part of the light cone that is closer in redshift to one of the snapshots than any other snapshot (with exceptions near the boundaries discussed below). The boundary between two redshift slices with snapshot redshifts $z^{(k)}$ and $z^{(k+1)}$ is thus a plane at comoving distance $w_{\mrm{U}}^{(k)}=w_{\mrm{L}}^{(k+1)}=w\left[\left(z^{(k)}+z^{(k+1)}\right)/2\right]$. In addition, $w_{\mrm{L}}^{(0)}=0$.
The particle data of the snapshot closest in redshift is then used to approximate the matter distribution in each of these slices. Fast box-intersection tests \citep{GottschalkLinManocha1996_OBBTreePaper} and the spatial-oct-tree storage order of the simulation are utilised to minimise reading of the particle data (which reduces run time by factors 5-10).

In the construction of the matter distribution in the light cone, special care is taken for the particles near the boundary of two slices.
In the simulation, particle concentrations representing dark matter halos of galaxies or clusters were identified by a friend-of-friend (FoF) group finding algorithm. Some of these halos are located on the slice boundaries with particles on either side.\footnote{
Approx. 0.5\% (5\%) of halos with virial masses \mbox{$M_{200}\geq 10^{12}h^{-1}\Msolar$} ($\geq 10^{15}h^{-1}\Msolar$) are affected by this procedure. Though not essential for cosmic shear simulations (test show a relative difference $\sim0.1\%$ for the shear power spectra), the proper treatment of halos near slice boundaries is important for group-galaxy lensing and strong lensing simulations.}
In order to avoid that such a halo is only partially included into a slice (and hence would be only partially projected onto a lens plane), a halo is either included as a whole if its central particle is inside the slice as defined by boundary planes, or completely excluded otherwise.

\begin{figure}[tb]
\centerline{\includegraphics[width=1\linewidth]{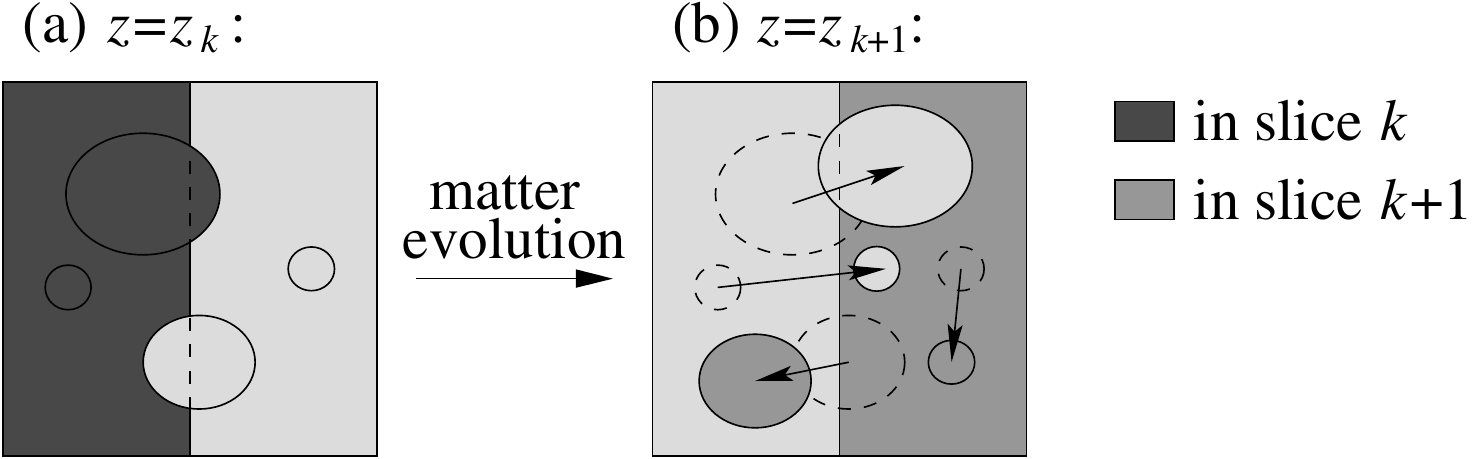}}
\caption{
\label{fig:rt_boundary}
Schematic view of the adaptive slice boundaries to avoid the truncation or double inclusion of halos that are located near a slice boundary. Halos near the boundary of slice $k$ and $k+1$ are either included as a whole in slice $k$ or completely excluded depending on the positions of their centres (a). Halos that are included (excluded) in slice $k$, are excluded (included) from slice $k+1$ even if they have crossed the slice boundary between redshift $k$ and $k+1$ (b).
}
\end{figure}

If the matter structure in the simulation were static, this procedure would suffice to prevent parts of the same halo from being projected onto adjacent lens planes, which would create artificial close pairs of halos on the sky. Halos, however, may have moved across a slice boundary between two snapshots. We therefore amend the above inclusion criterion for halos near the boundary: If a halo is included in (excluded from) the slice of the later snapshot, its progenitors in the earlier snapshot are excluded from (included in) the `earlier' slice even if their centres lie on the `early' ('late') side of the slice boundary. These inclusion criteria for halos are illustrated in Fig.~\ref{fig:rt_boundary}.

\subsection{The lens planes}
\label{sec:lens_planes}
\begin{figure}[tb]
\centerline{\includegraphics[width=1\linewidth]{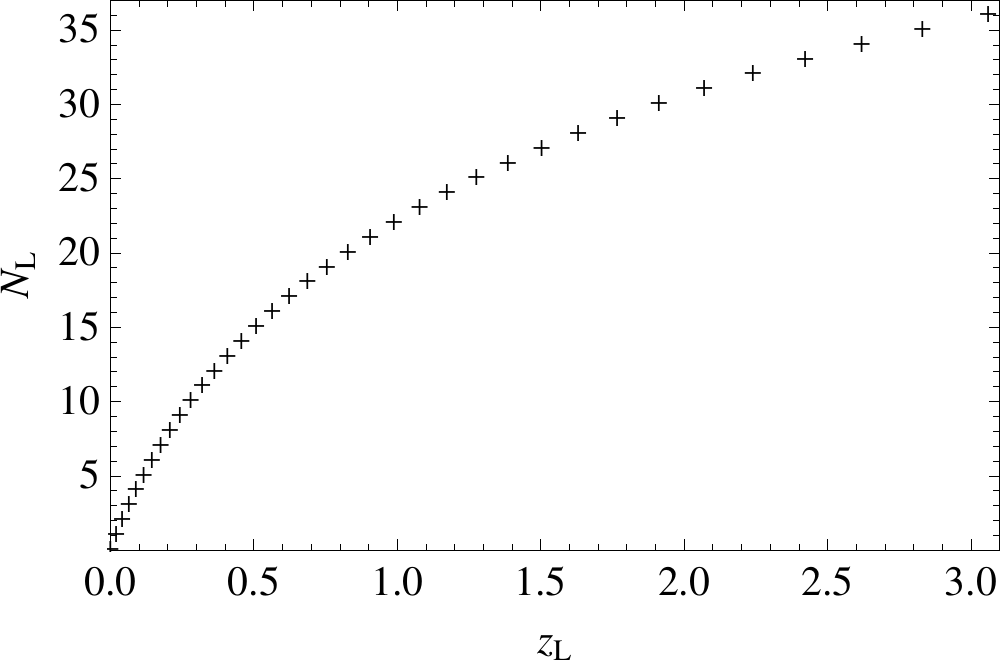}}
\caption{
\label{fig:n_planes_of_z}
The number $N_\mrm{L}$ of lens planes used for the ray-tracing as a function of the source redshift $z_\mrm{S}$.
}
\end{figure}

The matter content of each redshift slice of the backward light cone is projected along the LOS direction onto a lens plane. Each lens plane is placed at the comoving distance of the corresponding snapshot's redshift. The lens planes serve also as source planes for the ray-tracing. The resulting number of lens planes as a function of the source redshift is shown in Fig.~\ref{fig:n_planes_of_z}.

The light deflection angles and distortions resulting from the projected matter density on the lens planes are computed by particle-mesh (PM) methods \citep[][]{HockneyEastwood_book}. Mesh methods have the advantage that, once the deflection and distortion are computed on a mesh (e.g. by Fast Fourier methods), the computation of the deflections and distortions for many light rays intersecting the plane is very fast (compared to, e.g., direct-summation or tree methods). One disadvantage is that the used mesh spacing limits the spatial resolution of the projected matter distribution. However, any $N$-body simulation providing the matter distribution for the ray-tracing has a limited resolution as well.
In dense regions, the  spatial resolution of the Millennium Simulation is effectively determined by the force softening, which is $5h^{-1}\,\kpc$ comoving. Thus, a mesh spacing of $2.5h^{-1}\,\kpc$ comoving is required to avoid resolution degradation for the projected matter density.
However, a single mesh covering the full periodic area of the lens plane (i.e. $1.58h^{-1}\,\Gpc \times 1.66h^{-1}\,\Gpc$ comoving) with such a small mesh spacing would be too demanding, in particular regarding the memory required both for its computation and storage. We therefore use a hierarchy of meshes instead.

The lensing potential $\psi$ is split into a long-range part $\psi_\mrm{long}$ and a short-range part $\psi_\mrm{short}$. The split is defined in Fourier space by:
\begin{align}
\hat{\psi}_\mrm{long}(\vec{\ell})&=\hat{\psi}(\vec{\ell})\exp\left(-\beta_\mrm{split}^2\vec{\ell}^2\right)\mbox{, and}\\
\hat{\psi}_\mrm{short}(\vec{\ell})&=\hat{\psi}(\vec{\ell})\left[1-\exp\left(-\beta_\mrm{split}^2\vec{\ell}^2\right)\right].
\end{align}
The splitting angle $\beta_\mrm{split}=r_\mrm{split}/f_K(w)$, with comoving splitting length $r_\mrm{split}$ and comoving angular diameter distance of the lens plane $f_K(w)$, quantifies the spatial scale of the split. Different meshes are then used to calculate $\psi_\mrm{long}$ and $\psi_\mrm{short}$.

First, the particles in each slice are projected onto a coarse mesh of $16384\times16384$ points covering the whole periodic area of the lens plane using clouds-in-cells (CIC) assignment \citep{HockneyEastwood_book}. The long-range potential $\psi_\mrm{long}$ is then calculated on this mesh by means of fast Fourier transform (FFT) techniques~\citep{CooleyTukey1965_FFT,FrigoJohnson2005_FFTW3}. The splitting length $r_\mrm{split}=175h^{-1}\,\kpc$ is chosen slightly larger than the coarse mesh spacing ($96h^{-1}\,\kpc$ and $101h^{-1}\,\kpc$ comoving, respectively), so the coarse mesh samples $\psi_\mrm{long}$ with sufficient accuracy. For each lens plane, the long-range potential is calculated once, and the result is stored on disk for later use during the ray-tracing.

The short-range potential $\psi_\mrm{short}$ is calculated `on the fly', i.e. during the actual ray-tracing. The area where the light rays intersect the plane is determined and, if larger than $40h^{-1}\,\Mpc$ comoving, subdivided into several patches up to that size. Each patch is covered by a fine mesh with a mesh spacing of $2.5h^{-1}\,\kpc$ comoving and up to $16384\times16384$ mesh points. The fine meshes are chosen slightly larger than the patches in order to take into account all matter within the effective range of $\psi_\mrm{short}$, for which we assume $875 h^{-1}\,\kpc$ ($=5 r_\mrm{split}$). The limited range of $\psi_\mrm{short}$ ensures that the matter distribution outside the mesh affects only mesh points close to its boundary (i.e. within the effective range), but not the interior mesh points used for the subsequent analysis. Periodic boundary conditions can therefore be used for the FFT on the patches without `zero padding'.

In order to reduce the shot noise from the individual particles, either a fixed or an adaptive smoothing scheme is used for the matter distribution on the fine meshes. In case of the fixed smoothing, the particles in the slice are projected onto the fine mesh using CIC. The resulting matter density on the fine mesh is then smoothed in Fourier space with a Gaussian low-pass filter
\begin{equation}
\hat{K}_\mrm{s}(\vec{\ell};\beta_\mrm{s})=\exp\left(-\frac{\beta_\mrm{s}^2}{2}\vect{\ell}^2\right)
\end{equation}
whose filter scale $\beta_\mrm{s}=l_\mrm{s}/f_K(w)$ is determined by the lens plane's comoving distance $w$ and a fixed comoving filter scale $l_\mrm{s}$. This is done during the calculation of the short-range potential $\psi_\mrm{short}$ with FFT methods.

In case of the adaptive smoothing, the mass associated with each simulation particle contributes
\begin{equation}
 \Sigma_\mrm{p}(\vec{x})=
\begin{cases}
\dfrac{3 m_\mrm{p}}{\pi r_\mrm{p}^2}\left(1-\dfrac{|\vec{x}-\vec{x}_\mrm{p}|^2}{r_\mrm{p}^2}\right)^{2}, & |\vec{x}-\vec{x}_\mrm{p}|<r_\mrm{p},\\
 0, &  |\vec{x}-\vec{x}_\mrm{p}| \ge r_\mrm{p},
\end{cases}
\end{equation}
to the surface mass density on the fine mesh. Here, $\vec{x}$ denotes comoving position on the lens plane, $\vec{x}_\mrm{p}$ is the projected comoving particle position, and $r_\mrm{p}$ denotes the comoving distance to the 64th nearest neighbour particle in \emph{three} dimensions (i.e. before projection). The adaptive smoothing is essentially equivalent to the assumption that, in three-dimensional space, each simulation particle represents a spherical cloud with a Gaussian density profile and an {\it rms} radius that is half the distance to its 64th nearest neighbour. From the resulting surface mass density on the fine mesh, the short-range potential $\psi_\mrm{short}$ is then calculated by FFT methods.

The long- and short-range contributions to the deflection angles and shear matrices are calculated on the coarse and fine mesh by finite differencing of the potentials.\footnote{
The ray-tracing algorithm has to compute 5 derivatives of the lensing potential (2 deflection angle components and 3 shear matrix components) starting from the matter distribution. On large meshes, lower-order finite differencing operations (FDs) are much faster than FFTs. Using FFT derivatives would require 6 FFTs, whereas our approach requires only 2 FFTs and 5 FDs.
}
The values between mesh points are obtained by bilinear interpolation. The resulting deflection angles and shear matrices at the ray positions are then used to advance the rays and their associated distortion matrices from one plane to the next.

\begin{figure}[tb]
\centerline{\includegraphics[width=1\linewidth]{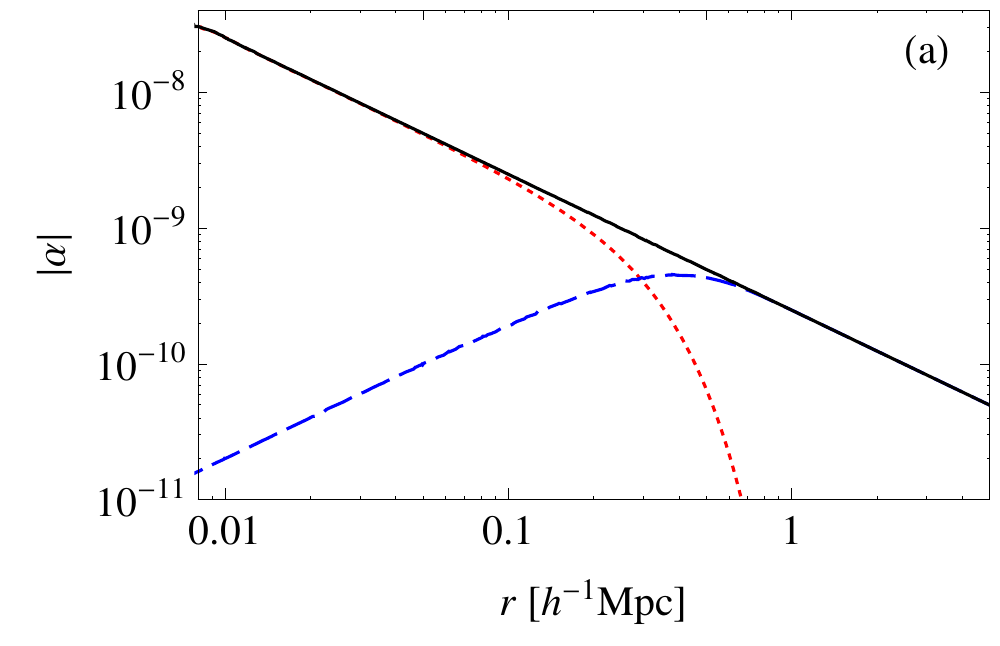}}
\centerline{\includegraphics[width=1\linewidth]{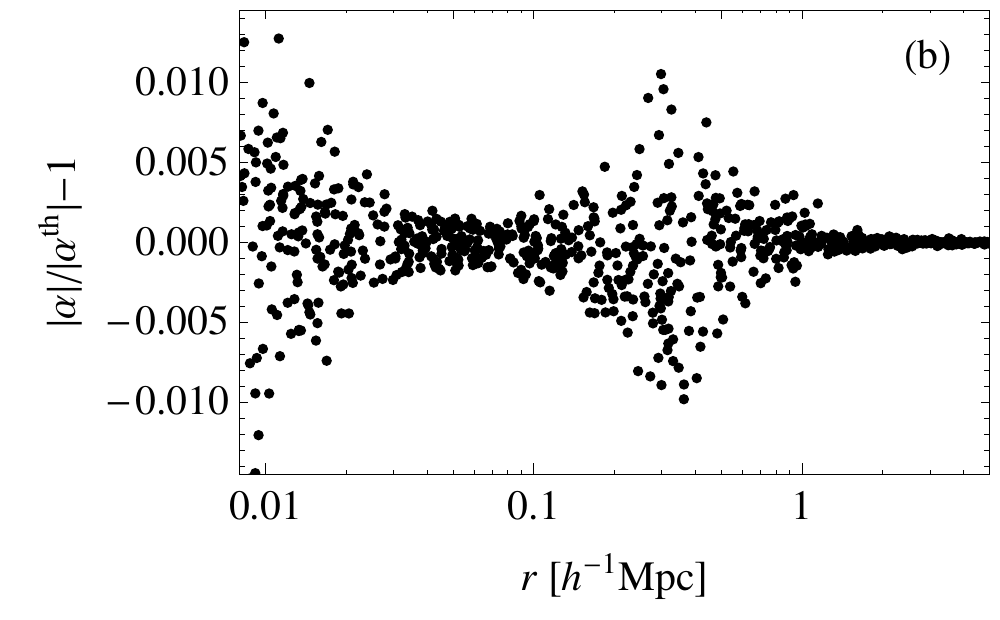}}
\caption{
\label{fig:point_mass_test_deflection}
The deflection angle $|\vec{\alpha}|$ as a function of the comoving distance $r$ to a point mass computed by our mesh method. (a) Short-range component (dashed line) and long-range component (dotted line) of the deflection angle (full line).
(b) Fractional difference between the value $\vec{\alpha}$ calculated by our mesh method and the theoretical value $\vec{\alpha}^\mrm{th}$. For the plots, we placed 10 point masses of $8.6\times10^8 h^{-1}\,\Msolar$ randomly on the lens plane at $z=0.5$ and calculated the resulting deflection at 1000 random positions around each of them.
}
\end{figure}

\begin{figure}[tb]
\centerline{\includegraphics[width=1\linewidth]{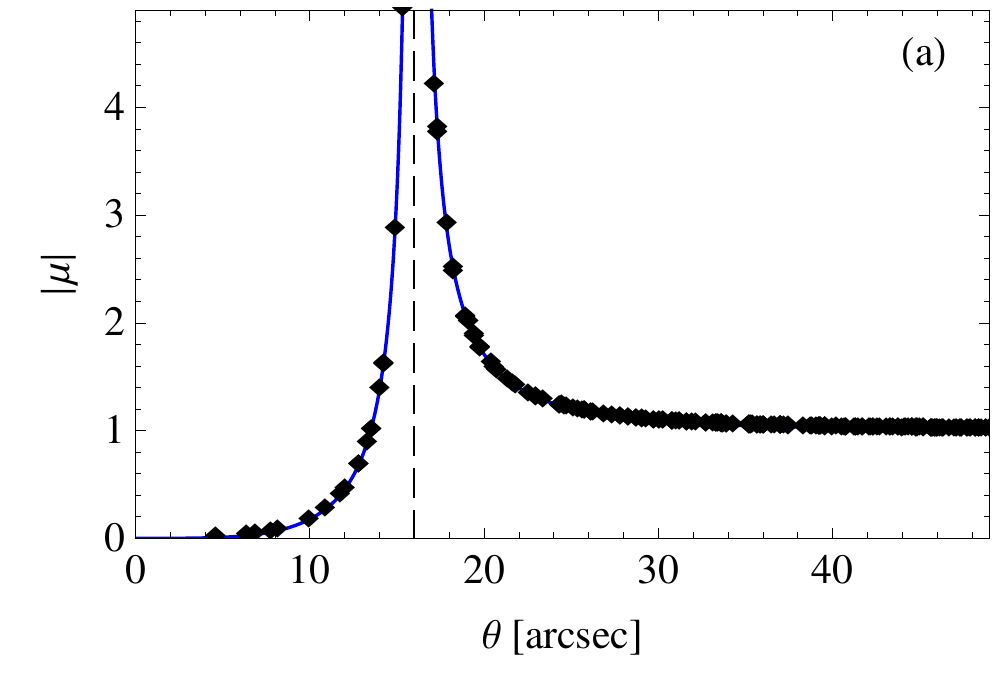}}
\centerline{\includegraphics[width=1\linewidth]{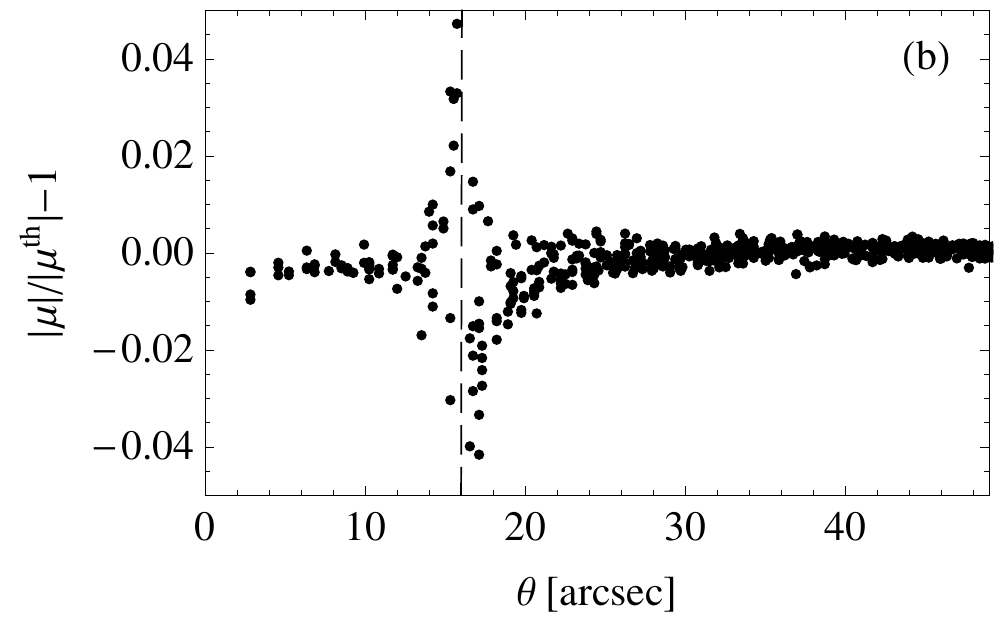}}
\caption{
\label{fig:point_mass_test_magnification}
The magnification $|\mu|$ for sources at $z=2$ as a function of the angular separation $\theta$ to a point mass of $10^{14}h^{-1}\,\Msolar$ at redshift $z=1$ computed by our mesh method. (a) Numerical values (symbols) compared to the analytical result (solid line). (b) Fractional difference between the measured magnification $|\mu|$ and its theoretical value $|\mu^\mrm{th}|$. The magnification diverges at the Einstein radius $\theta_\mrm{E}=16\arcsec$ (dashed vertical line). For the plots, we placed 10 point masses of $10^{14} h^{-1}\,\Msolar$ randomly on the lens plane at $z=1$ and calculated the resulting magnification at 1000 random positions around them.
}
\end{figure}

The ray-tracing algorithm reproduces the deflection angles and distortions caused by a single point mass very accurately, as is shown in Fig.~\ref{fig:point_mass_test_deflection} and \ref{fig:point_mass_test_magnification}. For the deflection angle, the relative deviation from the known analytical result is at most one percent, with a much smaller rms error. Apart from scales below $10h^{-1}\,\kpc$ comoving, where discreteness of the fine mesh becomes important, the largest relative errors occur on the scale of the split between short- and long-range potential. If desired, an even smaller error in this region could be obtained by increasing the splitting scale.

In this work, we do not consider the effects of the stellar mass in galaxies. Note, however, that the ray-tracing algorithm can be extended to include the gravitational effects of the stars in galaxies as described in \citet{HilbertEtal2008_StrongLensing_II}: The positions and stellar masses of the galaxies are inferred from semi-analytic galaxy-formation models implemented within the evolving dark-matter distribution of the Millennium Simulation \citep{SpringelEtal2005_Millennium,CrotonEtal2006,DeLuciaBlaizot2007}. The light deflection induced by the stellar matter is then calculated using analytic expressions for the projected stellar mass profiles on the lens planes. The error made by adding the stellar matter onto the lens planes, although the dark-matter particles of the simulation represent the \emph{total} matter, can then be compensated using extended analytic profiles with negative masses \citep[as was done, e.g., by][]{WambsganssOstrikerBode2008}.

\subsection{Lensing maps and image positions}
\label{sec:maps_and_galaxies}
To produce lensing maps, we set up rays starting from a fiducial observer on a regular grid in the image plane with an angular field size and resolution suitable for the particular application. The resulting shear and convergence maps may be used directly to obtain, e.g., the shear correlation functions and power spectra.

\begin{figure}[tb]
\centerline{\includegraphics[width=1\linewidth]{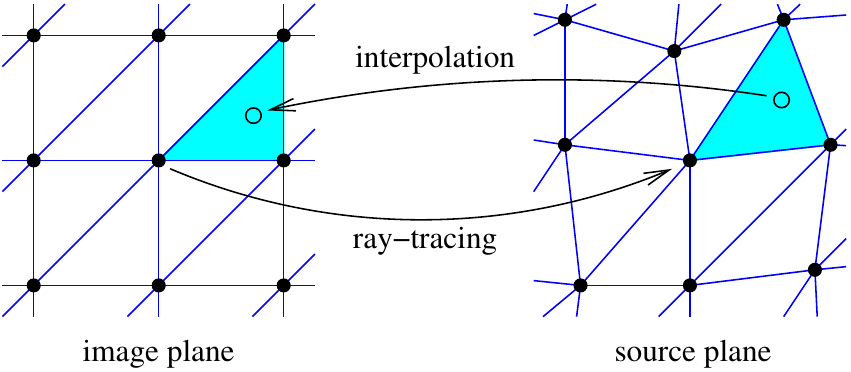}}
\caption{
\label{fig:triangle_interpolation}
Interpolation scheme used for determining image positions of galaxies. The regular grid of rays in the image plane (left filled circles) is used to partition the image plane into triangles (right blue lines). The image position (left open circle) of a source inside a triangle (right blue lines) formed by the backtraced rays on the source plane (right filled circles) is then determined by linear interpolation.
}
\end{figure}

We also wish to perform simulations of galaxy-galaxy lensing. Not only are the image positions of distant source galaxies affected by lensing. Also the apparent positions of galaxies and halos that act as lenses for background galaxies (and are to be correlated with the shear field) are affected by lensing due to foreground matter. We therefore have to compute the image positions $\vec{\theta}_{\rm g}$ given the galaxies' source positions $\vec{\beta}^{(k)}_{\rm g}$ (i.e. the projected galaxy positions on the lens planes) and the lens mapping sampled on the grid of light rays in the ray-tracing algorithm. To reach this, we make use of a triangle technique described, e.g., in \citet{SchneiderEhlersFalco_book}. We partition the region of the image plane that is covered by the grid of rays into triangles formed by rays of adjacent grid points (see Fig.~\ref{fig:triangle_interpolation}). On each lens plane, we identify for each such triangle all galaxies with source position inside the backtraced triangle. The image positions of these galaxies are then computed by linear interpolation between the rays.
This method takes into account strong lensing, as a galaxy might lie in more than one triangle on the lens plane, resulting in multiple images of that galaxy.

\begin{figure}[tb]
\centerline{\includegraphics[width=0.9\linewidth]{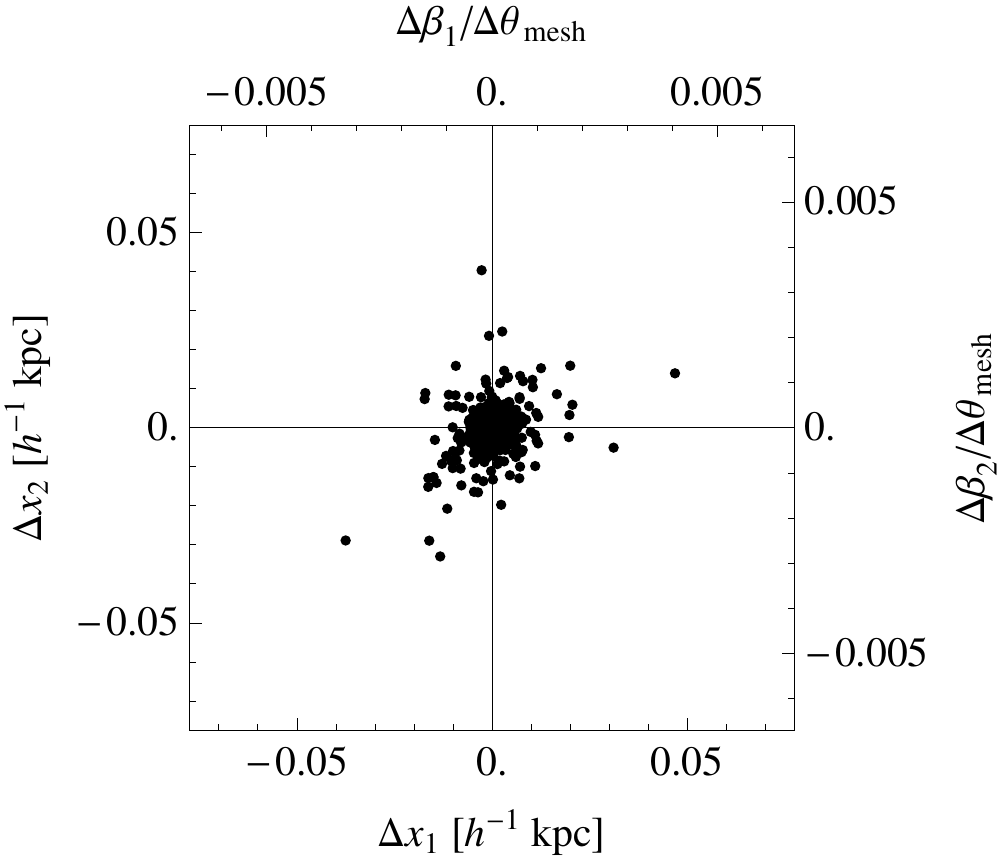}}
\caption{
\label{fig:interpolation_test}
Comparison of the true source positions and the source positions obtained by tracing back light rays through the Millennium Simulation starting from the image positions computed by the interpolation method. Shown is the difference $\Delta\vec{x}$ between true and traced-back comoving source position for $1\,000$ galaxy centres at $z=1$. At this redshift, $0.010h^{-1}\,\kpc$ comoving correspond to an angle of $10^{-3}\,\text{arcsec}$. The right and upper frames sides are labelled by the corresponding angular difference $\Delta\vec{\beta}$ between true and traced-back source position in units $\Delta\theta_\mrm{mesh}=1\arcsec$ of the mesh spacing used for the rays.
}
\end{figure}

Figure \ref{fig:interpolation_test} illustrates how well the image positions obtained by the triangle interpolation method are mapped back onto the source positions by the ray-tracing.Shown is the difference between the true source positions and the positions obtained by tracing back light rays starting from the interpolated image positions to the source plane. The difference is always much smaller than the resolution of the matter distribution on the lens planes. The slight anisotropy seen as a larger spread along one diagonal is caused by the particular way the image plane was partitioned into triangles. The diagonal coincides with the diagonal chosen for splitting the square mesh pixels into triangles. If one used the other diagonal for splitting the mesh pixels into triangles, a larger spread along that diagonal would be seen instead.

\section{Results}
\label{sec:results}
We compute various weak-lensing two-point statistics from a set of ray-tracing simulations and compare the results to semi-analytic predictions. If not stated otherwise, adaptive snoothing of the matter distribution on the lens planes is applied for the ray-tracing.

\subsection{Power spectra}
\label{sec:power_spectra}
We start our discussion with the convergence power spectrum $P_\kappa(\ell)$. In the first-order approximation \eqref{eq:cont_Jacobian_LinearApprox}, the convergence power spectrum is given by \citep[see, e.g.,][]{Schneider2006_SaasFee33_Part3}:
\begin{equation}
\label{eq:kps}
P_\kappa({\ell}) = \int_0^{\infty}{\rm d}w\,q^2(w)\,P_\delta\left(t(w),\frac{\ell}{f_K(w)}\right).
\end{equation}
Here,
\begin{equation}
\label{eq:geom_lensing_weight}
q(w)=\frac{3H_0^2\Omega_{\rm m}}{2c^2 a(w)} \int_w^{\infty} \diff{w'}\, p_\mrm{s}(w')\frac{f_K(w'-w)}{f_K(w')},
\end{equation}
with the probability distribution $p_\mrm{s}(w)$ of visible sources in comoving distance. Furthermore,  $P_\delta(t,k)$ denotes the three-dimensional power spectrum of the matter contrast $\delta$ at cosmic time $t$ and comoving wave vector $k$. For an accurate comparison with the results obtained by our ray-tracing algorithm, we use Eq.~\ref{eq:kps} together with the three-dimensional power spectra $P_\delta\left(k\right)$ measured from the Millennium Simulation \citep[see also][]{ValeWhite2003}. In the following, we will call the resulting power spectra \emph{first-order prediction} for brevity.

\begin{figure}[t]
\centerline{\includegraphics[width=\linewidth]{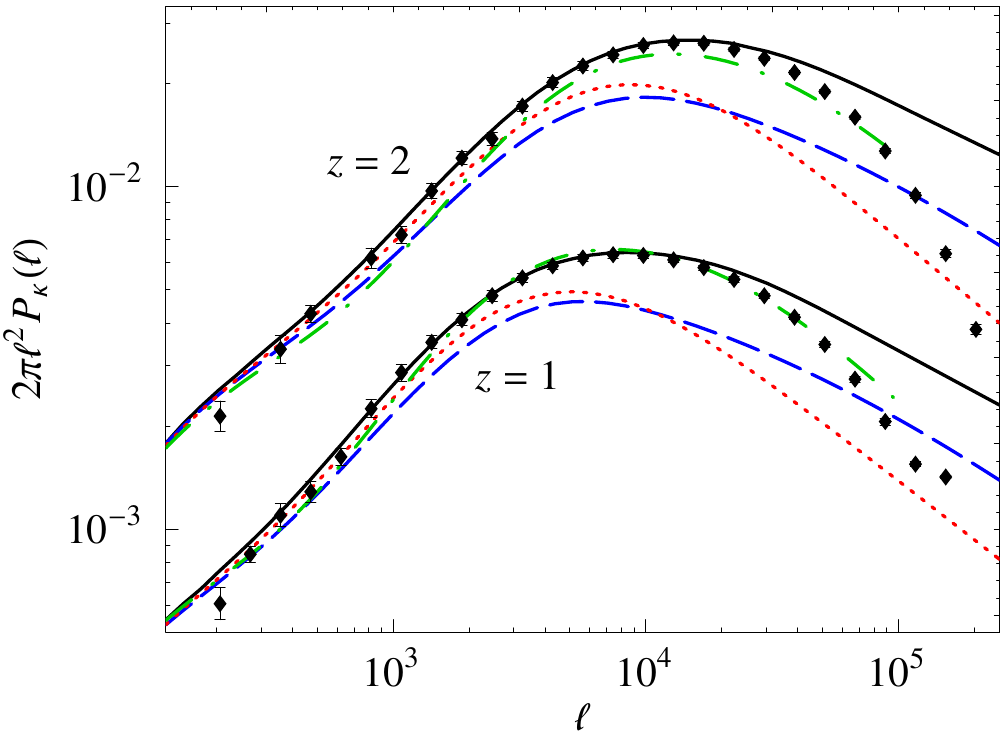}}
\caption{
\label{fig:power_spectra}
Convergence power spectra $P_\kappa(\ell)$ for sources at redshift $z=1$ (lower curves) and $z=2$ (upper curves). The simulation results (from $\sim30$ random fields of $3\times3\,\degt^2$) are shown as diamonds with errorbars (indicating standard deviation calculated from the field-to-field variance), the corresponding first-order predictions as solid lines. The predictions using the \cite{PeacockDodds1996} prescription together with the transfer function from \cite{EisensteinHu1999} are given as dotted lines, those obtained from the \cite{SmithEtal2003} fitting formula as dashed lines. The predictions of a halo model using the  concentration parameters of \citet{NetoEtal2007} are shown as dash-dotted lines. 
}
\end{figure}

In Fig.~\ref{fig:power_spectra}, we compare the first-order prediction to the convergence power spectra obtained from the ray-tracing. As has already been observed by \citet{JainSeljakWhite2000} and \citet{ValeWhite2003}, the power spectra from the ray-tracing are in very good agreement with the first-order prediction. For the considered source redshifts, the difference $\lesssim2\%$ for $\ell<10^4$. The larger deviations at wave numbers $\ell>2\times10^4$ are due to smoothing effects discussed below.

In the first-order approximation, the power spectra of the convergence and the shear are identical. As has already been found, e.g., by \citet{JainSeljakWhite2000}, the convergence and shear power spectra from ray-tracing agree very well, too. On scales $\ell>1000$, the difference between both is well below one percent in our ray-tracing results.

If the first-order prediction for the convergence power-spectra is assumed to be correct to very high accuracy, the smoothing tests can be considered as a test of the accuracy of our ray-tracing algorithm. Then the results shown in Fig.~\ref{fig:power_spectra} suggest that the ray-tracing is able to reproduce weak-lensing effects within $\sim3\%$ accuracy on scales  $300\lesssim\ell\lesssim20\,000$.

The comparison of the ray-tracing power spectra with some of the popular fitting formulae is less encouraging: Both the prescriptions by \cite{PeacockDodds1996} \citep[with the transfer function by][]{EisensteinHu1999} and \cite{SmithEtal2003} strongly underpredict the power on intermediate and small scales. These fitting formulae are based on older simulations, whose matter power spectra are noticeably different from the power spectra of more recent, higher-resolution simulations. The deviations from the simulated convergence power spectra exceed 30\% for $\ell>10000$, so these fitting formulae seem to be of limited use for the interpretation of data from future weak-lensing surveys. 

A prediction based on the popular halo model \citep{Seljak2000,CooraySheth2002} and the halo concentration-mass relation of \citet{NetoEtal2007} provides a better fit to the convergence power spectrum. There are, however, still deviations ($\approx 10\%$), in particular for higher source redshifts and intermediate scales (i.e. $\ell \approx 1-2\times 10^3$). This coincides with the transition region of the one- and two-halo terms, which is difficult to model accurately due to halo exclusion effects \citep[see e.g.~][and references therein]{TinkerEtal2005}, which are not included in our prediction.

\begin{figure}[t]
\centerline{\includegraphics[width=\linewidth]{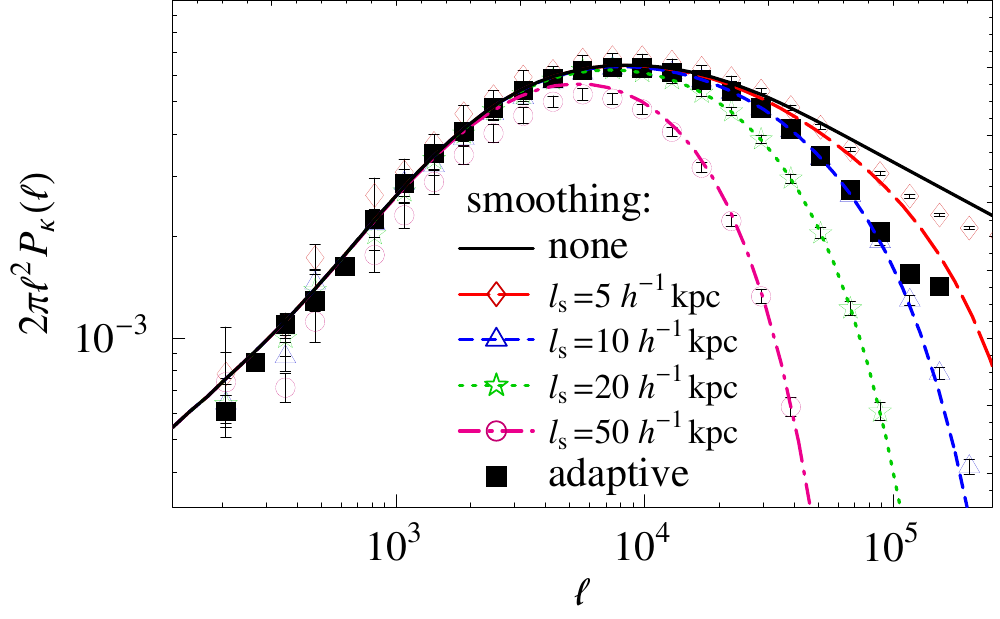}}
\caption{
\label{fig:ps_smooth}
Convergence power spectra $P_\kappa(\ell)$ for sources at redshift $z=1$. Compared are the results from ray-tracing (symbols) using various smoothing schemes (none/Gaussian with fixed scale $l_{\rm s}$/adaptive) and the corresponding first-order prediction (lines) obtained projecting and smoothing the measured 3D power spectra of the actual mass distribution in the simulation.
}
\end{figure}

As mentioned above, the deviations of the measured power spectra and the first-order predictions at large $\ell$ are due to smoothing effects.
In Fig.~\ref{fig:ps_smooth}, we present the convergence power spectra from ray-tracing runs of the same set of fields (with a cumulative area of $80\;{\rm deg}^2$ and sources at $z=1$), but with different smoothing schemes. In addition to adaptive smoothing, which is intractable analytically, we also employ smoothing with a Gaussian kernel of fixed comoving size on the lens planes. The ray-tracing simulations with Gaussian smoothing on the lens planes show -- apart from sampling variance -- perfect agreement with the first-order prediction if the smoothing is into taken into account there. Only the spectrum for the smallest smoothing length shows some aliasing effects on very small scales. The spectrum of the adaptive-smoothing runs happens to match the spectrum for a Gaussian smoothing length of $10h^{-1}\,\kpc$ comoving quite well, but one should be cautious when considering this as an "effective" smoothing length in a different context.

\subsection{Aperture-mass statistics}
A suitable cosmic-shear measure that allows one to decompose the shear signal in a finite-sized field into E- and B-modes is the aperture mass dispersion \citep{SchneiderEtal1998_Map,SchneiderVanWaerbekeMellier2002}. The E- and B-mode aperture mass at position $\vec{\theta}$ on the sky and scale $\vartheta$ are defined by:
\begin{equation}
\label{eq:mapdef}
M_{\rm E,B}^2 (\vect{\theta},\vartheta) = \int \diff[2]{\vec{\theta}'}\; Q\left(\vec{\theta}'-\vec{\theta},\vartheta\right)\,\gamma_{\mrm{t},\times}(\vec{\theta}',\vec{\theta}'-\vec{\theta}).
\end{equation}
In this work we use the polynomial filter function $Q$ proposed by \citet{SchneiderEtal1998_Map}:
\begin{equation}
Q\left(\vect{\theta},\vartheta\right)=
\frac{6|\vec{\theta}|^2}{\pi \vartheta^4} \left(1-\frac{|\vec{\theta}|^2}{\vartheta^2}\right)
.
\end{equation}
The tangential and cross components of the shear are defined by
\begin{subequations}
\begin{align}
\gamma_{\mrm{t}}(\vec{\theta}',\vec{\theta})=&
-\Re\left(\gamma(\vec{\theta}')\e^{-2\mrm{i}\phi(\vect{\theta})}\right)
,\\
\gamma_{\times}(\vec{\theta}',\vec{\theta})=&
-\Im\left(\gamma(\vec{\theta}')\e^{-2\mrm{i}\phi(\vect{\theta})}\right)
,
\end{align} 
\end{subequations}
where $\phi(\vect{\theta})$ is the polar angle for the direction defined by $\vect{\theta}$. 

\begin{figure}[t]
\centerline{\includegraphics[width=\linewidth]{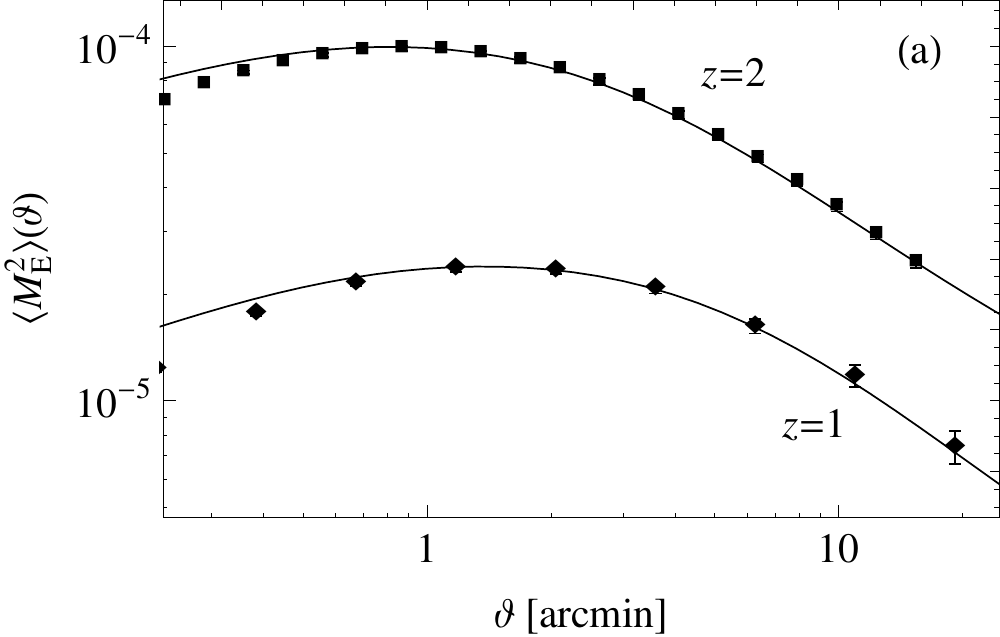}}
\centerline{\includegraphics[width=\linewidth]{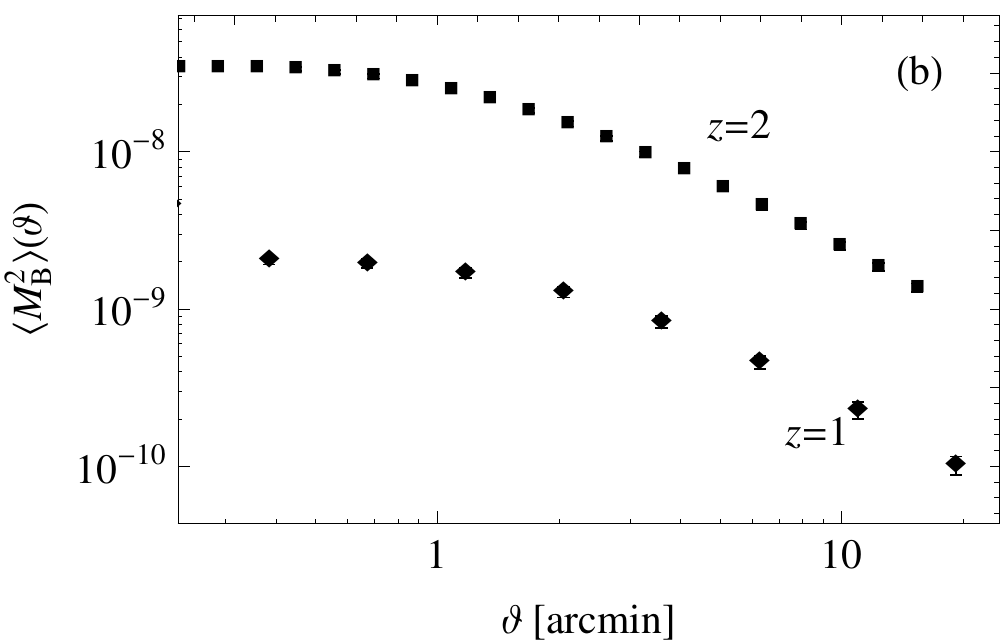}}
\caption{
\label{fig:mrmap}
Aperture mass dispersion $\EV{M_\mrm{E}^2}(\vartheta)$ (a) and  $\EV{M_\mrm{B}^2}(\vartheta)$ (b) as a function of filter scale $\vartheta$ for sources at $z=1$ and $z=2$. Compared are the first-order prediction (solid lines, E-mode only) and the results from ray-tracing (symbols with error bars indicating standard deviation, obtained from 7 fields of $5\times5\,\degt^2$).
}
\end{figure}
\begin{figure}[t] 
\centerline{\includegraphics[width=\linewidth]{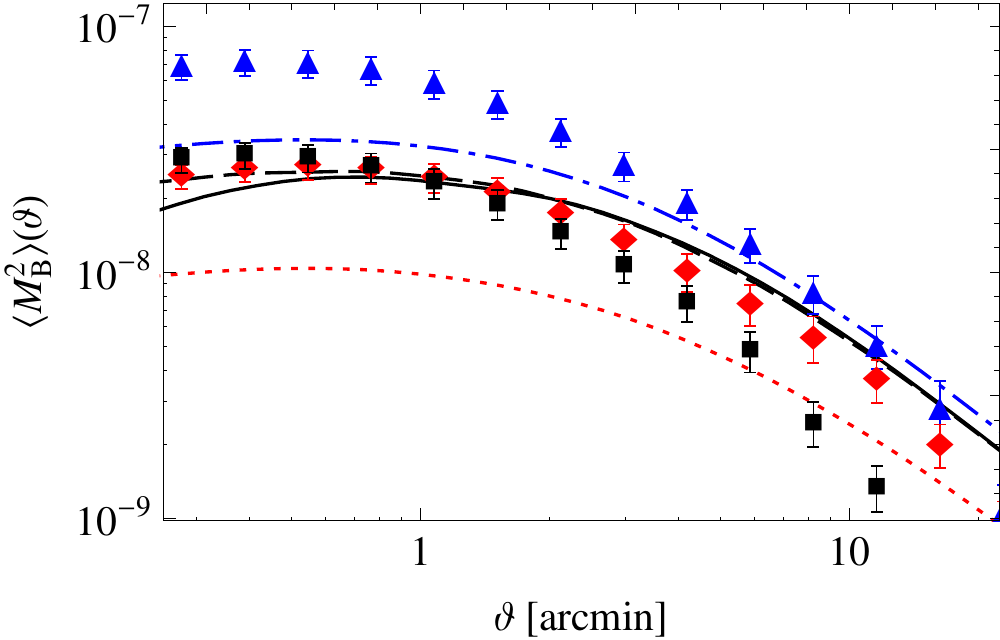}}
\caption{
\label{fig:bmodesth}
B-mode aperture mass dispersion $\EV{M_\mrm{B}^2}$ for sources at $z=2$ decomposed into the contributions by lens-lens coupling and Born corrections. Ray-tracing results (symbols with error bars indicating standard deviation, calculated from 7 fields of $1\times1\,\degt^2$): full ray-tracing (squares), only Born corrections (diamonds), only lens-lens coupling (triangles). Predictions by \cite{CoorayHu2002}: full signal (dashed line), only Born corrections (dotted line), only lens-lens coupling (dash-dotted line).
Prediction by \cite{HirataSeljak2003}: full signal (solid line).
}
\end{figure}

An estimate for the aperture mass dispersion  $\EV{M_{\rm E,B}^2}(\vartheta)$ as a function of the filter scale $\vartheta$ can be computed from a given shear field by a spatial average. Figure \ref{fig:mrmap}a shows the E-mode aperture mass dispersion measured from our set of simulations. The  dispersion measured from the ray-tracing is in very good agreement with the first-order prediction \citep{SchneiderEtal1998_Map}:
\begin{equation}
\label{eq:map_LinearApprox}
\EV{M_{\rm E}^2} (\vartheta) = \frac{288}{\pi}
 \int_0^\infty \diff{\ell} 
\frac{\ell\operatorname{J}_{4}^2(\vartheta\ell)}{(\vartheta\ell)^4}
 P_\kappa(\ell)
,
\end{equation}
where $P_\kappa(\ell)$ is given by Eq.~\eqref{eq:kps}, and $\operatorname{J}_{4}$ denotes a Bessel function of the first kind. The deviations of the measured aperture mass dispersion from the first-order prediction seen on scales $\lesssim0.5\,\arcmint$ can be attributed to smoothing.

In the first-order approximation, the B-mode aperture mass dispersion $\EV{M_{\rm B}^2} (\vartheta)$ vanishes. The measured B-mode dispersion from the full ray-tracing is shown in Fig.~\ref{fig:mrmap}b. The B-mode signal is at least 3 orders of magnitude smaller than the E-mode. On larger scales their ratio even drops below $10^{-5}$.

Theoretical predictions of the amplitude of the lensing-induced B-modes have been made by \cite{CoorayHu2002} and \cite{HirataSeljak2003}, who calculated corrections to the E- and B-mode shear power spectra by expanding Eq.~\eqref{eq:cont_Jacobian} to second order in the gravitational potential.
As Fig.~\ref{fig:bmodesth} illustrates, the predictions based on their methods (and the measured three-dimensional power spectra of the Millennium Simulation) are of the correct order of magnitude and reproduce some qualitative features of the ray-tracing simulations, but the match is far from being perfect. While the B-mode predictions are lower by a factor of $\approx 2$ on small scales, the signal measured from the ray-tracing declines much more quickly on larger scales. However, the discrepancies are not large enough to challenge the finding of \citet{ShapiroCooray2006} that the lensing-induced B-mode is unimportant even for an all-sky survey.

In order to determine their individual contributions to the total B-mode signal, we switch off ray deflections [i.e.\ we employ the Born approximation by setting $\vec{\theta}^{(k)}=\vec{\theta}\; \forall k$ in Eq.~\eqref{eq:ang_pos}] and/or lens-lens coupling [i.e.\ we set $\tens{A}^{(k-1)}=\tens{1}$ in the third term on the r.h.s.\ of Eq.~\eqref{eq:jacob_rec}]. Again the predictions and the measured signal differ by factors $\sim2$ on small scales, and the measured signal decreases much stronger with increasing scale.

We closely examined various steps involved in the calculations to exclude numerical artifacts as the reason for the discrepancy. The smoothing tests show that the smoothing we applied to the matter distribution on the lens planes can only account for deviations on scales $\lesssim 0.5\,\arcmint$. Examining the variance between the different ray-traced fields, we can exclude `cosmic variance' as a major source of the discrepancy.

The ray-tracing results did not change when different ways of estimating $M_\mrm{B}^2$ in real and Fourier space, as well as different methods of numerical integration for the theoretical curves were used. Furthermore, only a tiny B-mode (at least 6 orders of magnitudes smaller than the E-mode) remained, when both ray deflections and lens-lens coupling were switched off in the simulation (which is essentially equivalent to the first-order approximation). The origin of this tiny signal is found to be the interpolation of the Jacobian matrix between the grid points to obtain their values at the light ray positions. Sampling a B-mode-free, continuous shear field on a grid and subsequent interpolation yields again a continuous shear field. This, however, agrees exactly with the original field only at the grid points. Therefore, it may in general contain a small B-mode contribution, depending on the grid resolution and the interpolation scheme used.

\subsection{Galaxy-galaxy lensing}

We test the effect of Born corrections and lens-lens coupling on galaxy-galaxy lensing  (GGL) by producing a catalogue of unbiased mock galaxies. We achieve this by first drawing a number of simulation particles on each lens plane at random and using their positions as lens galaxy positions in the algorithm described in Sec.\ \ref{sec:maps_and_galaxies}. We then obtain a catalogue of source galaxies by randomly sampling positions in the image plane assuming a uniform image distribution over the field-of-view.
 
The GGL signal we are interested in is given by the mean tangential shear $\EV{\gamma_{\rm t}}(\vartheta)$ at the image positions of the source galaxies as a function of angular separation $\vartheta$ to the positions of the lens galaxies. In the simple case of unbiased galaxies considered here, the expected GGL signal can be computed in the first-order approximation by:
\begin{multline}
\label{eq:ggl_gamma_t}
\EV{\gamma_{\rm t}}(\vartheta)= \frac{1}{2\pi} \int\!\diff{w} \frac{p_\mathrm{l}(w)q(w)}{f_K(w)}
\\\times
\int\!\diff{\ell}\ell \operatorname{J}_2(\vartheta\ell)P_\delta\!\left(t(w),\frac{\ell}{f_K(w)}\right)
,
\end{multline}
where ${\rm J}_2$ is a Bessel function of the first kind, $p_\mathrm{l}(w)$ is the probability distribution of the lens galaxies' distances, the lensing weight $q(w)$ is given by Eq.~\eqref{eq:geom_lensing_weight}, and $P_\delta$ denotes again the 3D matter power spectrum. For simplicity, we will consider a volume-limited sample of lens galaxies with constant comoving density in the following.

\begin{figure}[t]
\centerline{\includegraphics[width=\linewidth]{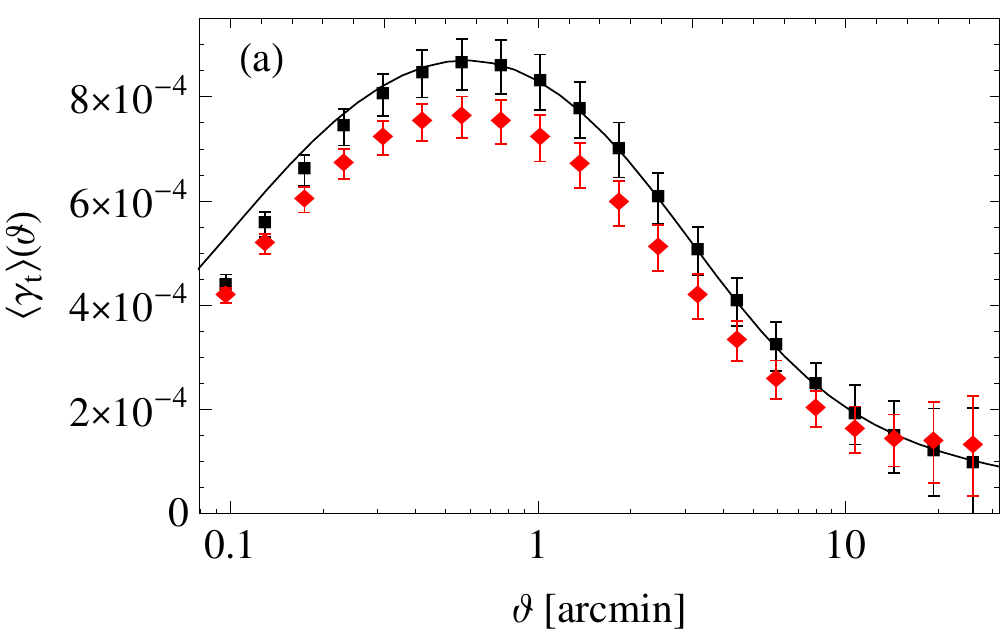}}
\centerline{\includegraphics[width=\linewidth]{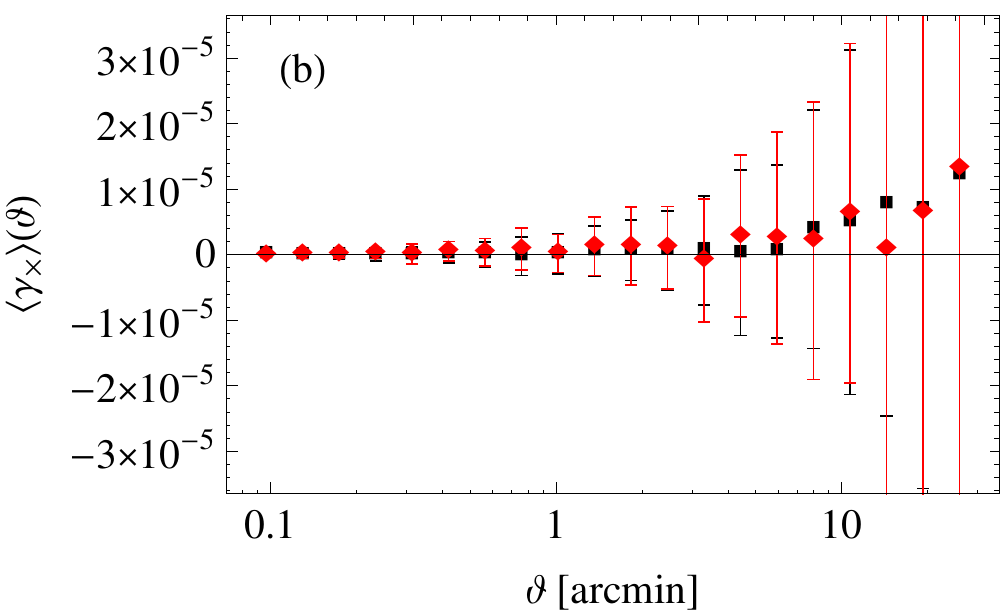}}
\caption{
\label{fig:GGL}
Galaxy-galaxy-lensing signal for sources at redshift $z=1$ and unbiased lens galaxies with a constant comoving mean density between $z=0$ and $z=1$. (a) Shown are the measured tangential component $\EV{\gamma_t}(\vartheta)$ of the shear from full ray-tracing (diamonds) and ray-tracing using the first-order approximation \eqref{eq:jacob_multi_plane_LinearApprox} (squares), and the first-order prediction \eqref{eq:ggl_gamma_t} (solid line). (b) Measured cross component $\EV{\gamma_\times}(\vartheta)$ from full ray-tracing (diamonds) and first-order ray-tracing (squares). Error bars denote the standard deviation calculated from a set of 24 simulated fields of $3\times3\,\degt^2$.
}
\end{figure}

Due to statistical parity invariance, the cross component $\gamma_{\times}$ is expected to vanish when averaged over many source-lens pairs. The observed mean cross component  $\left<\gamma_\times\right>$ can therefore be used as a test for systematic effects and `cosmic variance'. As shown in Fig.~\ref{fig:GGL}, $\left<\gamma_\times\right>$  is consistent with zero in our ray-tracing.

While the cross component $\gamma_\times$ provides a test for systematic effects, the tangential shear $\gamma_t$ contains the desired information about the matter and galaxy distribution. As can be seen in Fig.~\ref{fig:GGL}, the mean tangential shear $\left<\gamma_t\right>$ is significantly smaller ($\approx 10-20\%$ at an angular separation of $1\,\arcmint$) in the ray-tracing than expected from the first-order prediction \eqref{eq:ggl_gamma_t}.

The reason for this discrepancy is magnification bias: Lenses, i.e. dense matter structures such as galaxies or clusters with their dark matter halos, magnify the regions behind them. The magnification reduces the apparent number density of higher-redshift lens galaxies around lower-redshift lenses in a volume limited survey (as has been simulated here). Underdense regions, on the other hand, demagnify the regions behind them, thereby increasing the apparent number density of lens galaxies behind them. The de-/magnification leads to an anticorrelation between the positions of high-redshift lens galaxies and the tangential shear induced by low-redshift structures. The anticorrelation reduces the signal $\left<\gamma_t\right>$ compared to the first-order approximation.\footnote{Note that in the first-order approximation, magnification effects are neglected. Thus, the positions of galaxies at any given redshift are uncorrelated with the shear induced by galaxies at different redshifts.}
We can suppress the magnification bias in the ray-tracing by switching off the deflections and using Eq.~\eqref{eq:jacob_multi_plane_LinearApprox} to calculate the distortions. In this case our simulations are fully consistent with the first-order prediction, as is shown in  Fig.~\ref{fig:GGL}.

The effect of the magnification bias on the GGL depends on the redshift distribution of the sources and the lenses. Moreover, the shape of the lens luminosity function may be  important if the lens population is selected using a magnitude limit. For example, the first-order approximation may \emph{under}estimate $\left<\gamma_t\right>$ for a lens population with a very steep luminosity function near the survey magnitude limit. We reserve a more detailed investigation of this effect with realistic source and lens distributions for future work.

\section{Summary}
\label{sec:summary}

In this work, we have described a new variant of the multiple-lens-plane  algorithm, which is particularly suited for ray-tracing through very large cosmological $N$-body simulations. The algorithm differs in some important details from previous works. This allows us to take full advantage of the unprecedented statistical power offered by the large volume and high spatial and mass resolution of the Millennium Simulation. The features discussed include: a tilted line-of-sight (to avoid periodic repetition of structures along the line-of-sight), adaptive slice boundaries (to avoid the slicing and duplication of bound structures), adaptive smoothing of the projected matter distribution on the lens planes (to reduce shot noise from the particles), a mutliple-mesh method for calculating the light deflections and distortions at the lens planes (which takes into account the small-scale and large-scale structure simultaneously), and a  method to include galaxies (as lenses and sources) from semi-analytic galaxy-formation models in the ray-tracing process.

We have used the ray-tracing code and the Millennium Simulation to investigate the impact of lens-lens coupling and multiple ray deflections on various cosmic shear two-point statistics. We have computed convergence power spectra from a set of ray-tracing realisations. For testing and comparison, we have also computed a first-order prediction of the convergence power spectrum using the measured three-dimensional power spectra of the mass distribution in the Millennium Simulation. We find that this first-order prediction agrees very well with the ray-tracing results except for very small scales (the difference is $>5\%$ only for $\ell>20000$), where smoothing on the lens planes becomes important. 

Comparing the convergence power spectrum from the ray-tracing to the predictions based on the fitting formulae for the matter power spectrum by \citet{PeacockDodds1996} and \citet{SmithEtal2003}, we find significant discrepancies ($>30\%$ for $\ell>10000$), casting the usefulness of these fitting formulae for cosmological parameter estimation for future surveys into doubt. A prediction based on the popular halo model and the halo concentration-mass relation of \citet{NetoEtal2007} fits better, but there are still noticeable deviations, in particular for higher source redshifts ($\sim10\%$ for sources at $\zS=2$). This indicates a need for more accurate descriptions of matter power spectra.

Furthermore, we have computed the E- and B-mode aperture mass dispersion using our ray-tracing algorithm. We find the B-mode to be finite, but at least three orders of magnitude smaller than the E-mode. The amplitude of the B-mode is slightly larger and shows a different scale dependence than the predictions of \citet{CoorayHu2002} and \citet{HirataSeljak2003}. We have performed various tests to exclude numerical artifacts as the origin of the deviations. Despite these discrepancies, we can confirm the finding of \citet{ShapiroCooray2006} that the lensing-induced B-mode can be safely neglected even in an all-sky survey.

Corrections to the first-order approximation can have a considerable impact on galaxy-galaxy lensing. In the simple case of a volume-limited sample of unbiased lens galaxies and all sources at redshifts $z=1$, the first-order approximation overestimates the mean tangential shear around lenses by $\approx 10-20\%$ at an angular separation of $1\,\arcmint$ due to its failure to incorporate the magnification bias. The impact of the magnification bias on the galaxy-galaxy lensing signal depends on the survey selection criteria and the luminosity and redshift distribution of the sources and the lenses. A detailed investigation of this effect should be carried out in future work.

\begin{acknowledgements}
We thank Volker Springel, Jeremy Blaizot, Ole M{\"o}ller, Martin Kilbinger, Emilio Pastor-Mira, and Uro\v{s} Seljak for helpful discussions, Jasmin Pielorz for providing the halo model power spectra, and the referee for very useful comments. This work was supported by the DFG within the Priority Programme 1177 under the projects SCHN 342/6 and WH 6/3.
\end{acknowledgements}

\appendix

\section{Lattice planes}
\label{sec:appendix_lattice_planes}
The periodicity of our matter distribution along and perpendicular to the line-of-sight can be studied within the theory of crystal lattices \citep[see, e.g.,][]{AshcroftMermin_book}. Here, we give a practical explanation rather than a rigorous proof.

Consider an array of unit cubes forming a simple cubic lattice with lattice constant unity. Choose two linearly independent lattice vectors $\vec{p}$ and $\vec{q}$ with $\vec{p}=(p_1,p_2,p_3)$ and $\vec{q}=(q_1,q_2,q_3)$ and $p_i,q_i\in\Z$. These two vectors span a plane which is perpendicular to the lattice vector $\vec{n}$ with $\vec{n}=(n_1,n_2,n_3)=(p_1,p_2,p_3)\times(q_1,q_2,q_3)$,  $n_i\in\Z$.

Since the plane-spanning vectors are lattice vectors, the plane is itself periodic and therefore represents a plane lattice. With $\vec{p}$ and $\vec{q}$ as basis vectors of the plane lattice, the plane is periodic along the direction of $\vec{p}$ and $\vec{q}$ with periodicity length $|\vec{p}|$ and $|\vec{q}|$, respectively. The parallelogram constructed from $\vec{p}$ and  $\vec{q}$ represents a unit cell of the plane lattice with a cell area of $|\vec{p}\times\vec{q}|=|\vec{n}|$. One can show that there is no smaller unit cell if the integer coefficients $n_1$, $n_2$, and $n_3$ are coprime. Furthermore, there is no shorter non-zero lattice vector perpendicular to the plane than $\vec{n}$ in this case, and hence, the shortest periodicity along the normal direction is $|\vec{n}|$.

For the computational cube of the Millennium Simulation with side length $L=500h^{-1}\,\Mpc$, the lengths and areas above have to be multiplied by $L$ and $L^2$, respectively. Our choices $\vec{p}=L(3,-1,0)$ and $\vec{q}=L(1,3,-1)$ yield a LOS vector $\vec{n}=L(1,3,10)$ with $|n|=5.244h^{-1}\,\Gpc$ and a rectangular area of $1.581h^{-1}\,\Gpc \times  1.658h^{-1}\,\Gpc$ for the lens planes.

\bibliographystyle{aa}


\end{document}